\documentstyle[12pt]{article}

\def\ar{\rightarrow}
\def\bib{\bibitem}

\def\dsl{\partial\!\!\!/}

\def\intx{\int\! d^{\sl 4}x}
\def\inty{\int\! d^{\sl 4}y}
\def\intX{\int\! d^{\sl 4}X\,}
\def\intY{\int\! d^{\sl 4}Y\,}
\def\intkm{\int\! \frac{d^{\sl 3}k}{(2{\pi})^3}\frac{m}{k^{\sl 0}}}
\def\intko{\int\! \frac{d^{\sl 3}k}{(2{\pi})^3\, 2\om_k}}
\def\intk{\int\! \frac{d^{\sl 4}k}{(2{\pi})^4}}
\def\intK{\int\! \frac{d^{\sl 4}K}{(2{\pi})^4}\,}

\def\ksl{k\!\!\!/}
\def\lar{\longrightarrow}

\def\midd{\! \mid \!}

\def\pa{\partial}
\def\plangle{{^P\! \langle}}
\def\prangle{{\rangle \!^P}}

\def\rvec{\!\!\!\!^{^\rightarrow}}
\def\lvec{\!\!\!\!^{^\leftarrow}}
\def\lrvec{\!\!\!\!\!\!^{^\leftrightarrow}}

\def\Vreg{V\!\!\!\!\!^{_{reg}}}

\def\al{\alpha}
\def\be{\beta}
\def\ga{\gamma}
\def\de{\delta}
\def\ep{\varepsilon}

\def\la{\lambda}
\def\va{\varphi}
\def\si{\sigma}
\def\om{\omega}

\def\Del{{\it\Delta}}
\def\Ep{{\cal E}}
\def\Ga{{\it\Gamma}}
\def\La{{\it\Lambda}}

\def\Pit{{\it\Pi}}

\def\Th{{\it\Theta}}

\def\beq{\begin{equation}}
\def\eeq{\end{equation}}
\def\bed{\begin{displaymath}}
\def\eed{\end{displaymath}}
\def\beqq{\begin{eqnarray}}
\def\eeqq{\end{eqnarray}}
\def\bedd{\begin{eqnarray*}}
\def\eedd{\end{eqnarray*}}

\def\ss{\scriptsize}

\begin{document}

\centerline{\normalsize\bf ASYMPTOTIC STATES AND THE DEFINITION OF} \centerline{\normalsize\bf THE $S$-MATRIX IN QUANTUM GRAVITY}

\vspace*{0.9cm}
\centerline{\footnotesize C. WIESENDANGER}
\baselineskip=12pt
\centerline{\footnotesize\it Aurorastr. 24, CH-8032 Zurich}
\centerline{\footnotesize E-mail: christian.wiesendanger@ubs.com}

\vspace*{0.9cm}
\baselineskip=13pt
\abstract{Viewing gravitational energy-momentum $p_G^\mu$ as equal by observation, but different in essence from inertial energy-momentum $p_I^\mu$ naturally leads to the gauge theory of volume-preserving diffeormorphisms of an inner Minkowski space ${\bf M\/}^{\sl 4}$. The generalized asymptotic free scalar, Dirac and gauge fields in that theory are canonically quantized, the Fock spaces of stationary states are constructed and the gravitational limit - mapping the gravitational energy-momentum onto the inertial energy-momentum to account for their observed equality - is introduced. Next the $S$-matrix in quantum gravity is defined as the gravitational limit of the transition amplitudes of asymptotic $in$- to $out$-states in the gauge theory of volume-preserving diffeormorphisms. The so defined $S$-matrix relates $in$- and $out$-states of observable particles carrying gravitational equal to inertial energy-momentum. Finally generalized LSZ reduction formulae for scalar, Dirac and gauge fields are established which allow to express $S$-matrix elements as the gravitational limit of truncated Fourier-transformed vacuum expectation values of time-ordered products of field operators of the interacting theory. Together with the generating functional of the latter established in \cite{chw4} any transition amplitude can in principle be computed consistently to any order in perturbative quantum gravity.}

\normalsize\baselineskip=15pt

\section{Introduction}

The Standard Model of the electromagnetic, weak and strong interactions is based on connecting observed conservation laws through Noether's theorem with global symmetries of matter field theories which then are made local requiring the existence of gauge fields in the process.

More specifically field theory allows to link conservation laws for e.g. the electric charge, weak iso-spin, colour, energy- and angular-momentum to specific symmetries of the action because fields provide the necessary inner and space-time degrees of freedom to carry representations of symmetry groups and because the Lagrangian approach allows to easily construct actions invariant under these symmetry transformations. The Noether mechanism then guarantees for each symmetry the existence of a conserved current and charge which can be interpreted in terms of the observed conserved quantity. This has been successfully applied to the conservation of e.g. the electric charge requiring a field theory to be invariant under $U(1)$, to the conservation of weak iso-spin requiring invariance under $SU(2)$, the conservation of colour requiring invariance under $SU(3)$ and the conservation of energy- and angular-momentum requiring invariance under the Poincar\'e group.

By construction the actions one starts with are only invariant under global symmetry transformations, but not under local ones. The introduction of a covariant derivative with a gauge field living in the adjoint representation of the symmetry group ensures the invariance of the action also under local transformations. Assuming renormalizable matter Lagrangians, minimal coupling - replacing ordinary by covariant derivatives - preserves the renormalizability of the matter Lagrangian and requiring a renormalizable gauge field Lagrangian uniquely determines the latter by dimensional analysis. This program applied to the inner symmetries $U(1), SU(2), SU(3)$ successfully leads to the photon field/Electrodynamics, the electro-weak gauge bosons/EW interaction, the gluon field/Chromodynamics \cite{lor,cli,stw1,stw2}. And - dropping the renormalizability condition - gauging the translation group as a space-time symmetry leads to general relativity (GR) or generalizations thereof \cite{chw1}.

But can the above line of thinking also be fruitful in suggesting alternative theories of gravitation? The original line of thinking in developing GR starts with the observation that inertial and gravitational mass have the same experimental values. The equivalence principle then postulates their essential identity which necessarily leads to a geometrization of gravity and to GR.

We propose to follow another route following closely the above gauge program \cite{chwA,chwB,chw2,chw3,chw4}: let us keep inertial and gravitational mass (or more appropriately inertial energy-momentum and gravitational energy-momentum) as two essentially different entities in the interacting theory, thereby postulating the existence of two different conserved four-vectors - one related by Noether's theorem to space-time translation invariance and the second one to a new inner symmetry.

This inner symmetry has to generate a conserved four-vector serving as gravitational energy-momentum. Following the general program outlined above the most simple way to generate such a conserved four-vector is to have fields carry a representation of an inner global translation group. Invariance of the action then generates the required vector.

Below we will summarize the construction of the gauge field theory of the volume-preserving diffeomorphism group where the latter acts as an inner symmetry group \cite{chw2}. To uncover the physical content of the quantized theory we discuss in this paper its asymptotic states obtained by canonical quantization for both matter and gauge fields.

Linking the theory to observable quantities e.g. scattering amplitudes requires the numerical identification of gravitational with inertial energy-momentum through a limiting process or the introduction of a "gravitational limit". On the basis of this limit we then define the $S$-matrix in quantum gravity and finally establish the necessary generalizations of the usual LSZ reduction formulae for both matter and gauge fields which link observable gravitational scattering amplitudes to the Green functions of the interacting theory. The generating functional of the latter we have constructed in \cite{chw4} and also hinted there at the theory's renormalizability a full proof of which we have established in the meantime \cite{chw5}.

All together this defines a renormalizable theory which is technically complete and allows us to perturbatively compute any quantum scattering process of interest. As an application showing how the approach works in practice we have calculated in \cite{chw6} the gravitational scattering amplitude and cross-section of two Dirac particles to lowest order in perturbation theory.

\section{Gauge Theory of the Group of Volume-Preserving Diffeomorphisms}
In this section we present the consequences of viewing gravitational energy-momentum $p_G^\mu$ as equal by observation, but different in essence from inertial energy-momentum $p_I^\mu$ and how the conservation of both quantities can be accounted for in a field theory context leading to a gauge theory of the group of volume-preserving diffeomorphisms as outlined in \cite{chw2}.

To explore an alternative route to a viable field theory of gravitation let us go back to the very fundament - namely the two experimental observations that (1) the inertial and gravitational masses of a physical body are numerically equal, $m_I = m_G$, and that (2) the inertial energy-momentum of a closed physical system is conserved, $p_I^\mu = conserved$. Taking both together we then can write in the rest frame of the body
\beq \label{0}
p_I^\mu = (m_I,\underline{0}) =^{\!\!\!\!\!\!\! ^{ (1)}} (m_G,\underline{0}) = p_G^\mu,
\eeq
where we have tentatively introduced the gravitational energy-momentum $p_G^\mu$ which we keep as an entity a priorily different from $p_I^\mu$. Note that $p_G^\mu$ is conserved due to Eqn.(\ref{0}).

In GR $m_I = m_G$ is interpreted as an essential identity which leads to the usual geometrical description of gravitation.

As discussed in detail in \cite{chw2} we propose to follow a different route and investigate the consequences of viewing $m_I$ and $m_G$ or $p_I^\mu$ and $p_G^\mu$ as different by their very natures - the prevailing view before Einstein which comes at the price of accepting the observed numerical equality $m_I = m_G$ as accidential.

Both $p_I^\mu$ and $p_G^\mu$ are four-vectors which are conserved, but in our approach through two different mechanisms. Obviously the conservation of $p_I^\mu$ is related to translation invariance in space-time. Let us use Noether's theorem to separately derive the conservation of a new four-vector in a field theoretical framework relating it to an inner continous symmetry of the theory specified below as invariance under volume-preserving diffeomorphisms. That four-vector is then interpreted as the gravitational energy-momentum $p_G^\mu$. 

To mathematically implement the idea we start with infinitesimal diffeomorphisms 
\beq \label{1}
X^\al\lar X'^\al = X^\al + {\cal E}^\al (X),\:\:\al=\sl{0,1,2,3}
\eeq
acting on ${\bf R}^{\sl 4}$ with points labelled by $X^\al$. $X'^\al (X)$ denotes an infinitesimal invertible and differentiable coordinate transformation of ${\bf R\/}^{\sl{4}}$ and $DIFF\,{\bf R}^{\sl 4}$ acts as a group on this space under composition.

To represent this group on fields we have to add the necessary degrees of freedom to the fields in question. So let us take a general matter field $\phi$ which has to be defined for our purposes on the product of the usual four-dimensional Minkowski space-time ${\bf M}^{\sl 4}$ and the four-dimensional inner space ${\bf R}^{\sl 4}$ introduced above
\beq \label{2}
\phi(x) \ar \phi(x,X)
\eeq
with Lagrangian
\beq \label{3}
\quad L_M = \intX \La^{-4} {\cal L}_M (\phi, \pa_\mu \phi). \nonumber
\eeq
Here, we have introduced a length scale $\La$ which is necessary to define an appropriate summation over the field components continuosly labelled by $X$. This summation measure has to be dimensionless as is the summation over the field components discretely labelled in the usual case of compact gauge groups - hence the necessity to introduce a length scale $\La$ so that $\intX \La^{-4}$ is dimensionless.

We note here that $\La$ plays a crucial role at different places in our approach and the theory - to be viable - cannot depend on the specific choice of such a $\La$. Hence, we have proven the equivalence of theories with different $\La$ based on an inner scale invariance of the theory established in Eqn.(27) in \cite{chw2} – which means that $\La$ can be chosen to be the Planck scale. This scale invariance is respected both by the regularization of inner volume integrals which occur at various places in the definition of the theory \cite{chw2,chw4} and by the quantized theory as the symmetry is linearly realized and hence survives quantization \cite{chw4}.

To next represent the group $DIFF\,{\bf R}^{\sl 4}$ on the matter field we take the passive view of diffeomorphisms
\beqq \label{4}
x^\mu &\lar& x'^\mu = x^\mu, \quad
X^\al \lar X'^\al = X^\al, \\
\phi (x,X) &\lar& \phi'(x,X) = \phi(x,X) -\, {\cal E}^\al (X)\cdot \nabla_\al\,\phi(x,X) \nonumber
\eeqq
transforming fields, but not coordinates.

Defining the variation $\de_{_{\cal E}} ..\equiv ..' - ..$ of an expression under an infinitesimal diffeomorphism or gauge transformation we find
\beq \label{5}
\de_{_{\cal E}} L_M = - \intX \La^{-4}
\,\nabla_\al \cdot \Big({\cal E}^\al (X) {\cal L}_M (\phi, \pa_\mu \phi)\Big)
= 0
\eeq
provided that the infinitesimal gauge parameter ${\cal E}^\al (X)$ is divergence-free
\beq \label{6}
\nabla_\al {\cal E}^\al (X) = 0. \nonumber
\eeq

This condition reduces the gauge group to volume-preserving diffeomorphisms ${\overline{DIFF}}\,{\bf R}^{\sl 4}$ which we will work with in the sequel. It is easily shown to be a group and restricts the ${\cal E}^\al (X)$ to be divergence-free.

For the Lagrangian Eqn.(\ref{3}) Noether's theorem then yields for any solution of the field equation four conserved currents
\beqq \label{7}
J^\nu\,_\al (x) &\equiv& \intX \La^{-4}\, \frac{\pa{\cal L}_M}{\pa(\pa_\nu \phi)} \nabla_\al\, \phi \\
\pa_\nu J^\nu\,_\al (x) &=& 0,\:\:\al=\sl{0,1,2,3} \nonumber
\eeqq
and four time-independent charges
\beq \label{8}
{\bf P}_\al\equiv \int\! d^{\sl 3}x \, J^{\sl 0}\,_\al, \:\:\al=\sl{0,1,2,3}
\eeq
which is the looked-for conserved inner four-vector generating global inner coordinate transformations in field space. After gauging the current Eqn.(\ref{7}) will serve as the source of the corresponding gauge field. Note that the above construction is transferable to any other given matter field.

For clarity a comparison with the usual Yang-Mills situation might help. In our approach the theory being invariant under volume-preserving diffeomorphisms comes along with gravitational momentum conservation -  accounted for by field theory means and Noether's law in analogy to how the conservation of e.g. colour in QCD is accounted for through an SU(3) global symmetry. In our case, however, the fields live on an infinite-dimensional space labelled by a continous index $X$ and not as in the QCD case on a three-dimensional vector space labelled by a discrete index. In \cite{chwA} we have worked out this analogy in more detail. Hence the theory is not eight-dimensional, but features four space-time dimensions which is also underlined by the four-dimensional field equations for the various fields displayed in Sections 3 and 4. In addition the fields carry a representation of the volume-preserving diffeomorphism group which is related to the four inner space dimensions. The global inner symmetry yields through Noether's theorem the conservation of gravitational momentum, gauging it leads to the gauge fields to which we turn next.

To obtain these we next allow for $x$-dependent infinitesimal volume-preserving gauge parameters ${\cal E}^\al (X)\ar {\cal E}^\al (x,X)$, $\nabla_\al {\cal E}^\al (x,X) = 0$. Note that Eqns.(\ref{4}) still define the representation of the gauge group in field space.

To ensure the local gauge invariance of the Lagrangian Eqn.(\ref{3}) we have to replace ordinary derivatives by covariant ones $\pa_\mu\ar D_\mu$
\beq \label{9}
D_\mu (x,X) \equiv \pa_\mu + A_\mu\,^\al (x,X)\cdot \nabla_\al
\eeq
thereby introducing gauge fields $A_\mu\,^\al (x,X)$ which are divergence-free
\beq \label{10}
\nabla_\al A_\mu\,^\al (x,X) = 0
\eeq
consistent with $\nabla_\be {\cal E}^\be (x,X)=0$. The transformation law for the gauge field is easily derived
\beq \label{11}
\de_{_{\cal E}} A_\mu\,^\al = \pa_\mu {\cal E}^\al + A_\mu\,^\be \cdot \nabla_\be {\cal E}^\al - {\cal E}^\be \cdot \nabla_\be A_\mu\,^\al 
\eeq
and respects $\nabla_\al \de_{_{\cal E}} A_\mu\,^\al = 0$.

By construction we find the Lagrangian Eqn.(\ref{3})
\beq \label{12}
\de_{_{\cal E}} L_M(\phi, D_\mu \phi)
= - \intX \La^{-4} \nabla_\al \Big({\cal E}^\al (x,X)\cdot {\cal L}_M (\phi, D_\mu \phi) \Big)
= 0
\eeq
to be invariant under local gauge transformations.

The field strength components are defined as usual by
\beq \label{13}
\left[D_\mu (x,X), D_\nu (x,X) \right] \equiv
F_{\mu\nu}\,^\al (x,X)\cdot \nabla_\al.
\eeq
They are expressed by the gauge fields as
\beqq \label{14}
& & F_{\mu\nu}\,^\al(x,X) = \pa_\mu A_\nu\,^\al(x,X) - \pa_\nu A_\mu\,^\al(x,X) \\ & &\quad + A_\mu\,^\be(x,X) \cdot \nabla_\be A_\nu\,^\al(x,X) - A_\nu\,^\be(x,X) \cdot \nabla_\be A_\mu\,^\al(x,X) \nonumber
\eeqq
and transform covariantly under a local gauge transformation
\beq \label{15}
\de_{_{\cal E}} F_{\mu\nu}\,^\al = F_{\mu\nu}\,^\be \cdot \nabla_\be {\cal E}^\al - {\cal E}^\be \cdot \nabla_\be F_{\mu\nu}\,^\al. 
\eeq

Finally, a gauge-invariant, minimal and renormalizable Lagrangian for the gauge field is given by \cite{chw2}
\beqq \label{16}
& & {\cal L}_G (A_\mu\,^\al, \pa^\nu A_\mu\,^\al; \nabla_\be A_\mu\,^\al) = \frac{1}{4\, \La^2}\, F_{\mu\nu}\,^\al(x,X) \cdot F^{\mu\nu}\,_\al(x,X), \\
& &\quad\quad\quad L_G = \intX \La^{-4} {\cal L}_G (A_\mu\,^\al, \pa^\nu A_\mu\,^\al; \nabla_\be A_\mu\,^\al). \nonumber
\eeqq
The Lagrangian involves contraction of inner indices by an inner metric \cite{chw2} which we have chosen to be the Minkowski metric $\eta_{\al\be}$ in inner space. Correspondingly we have partially fixed the gauge to those coordinate transformations which leave $\eta_{\al\be}$ invariant \cite{chw2}. In the sequel we will only work in these so-called Minkowskian gauges. As a result the inner space ${\bf R\/}^{\sl{4}}$ becomes a metric space $({\bf R\/}^{\sl{4}}, \eta)$ or inner Minkowski space ${\bf M}^{\sl 4}$ and points in that space transform as vectors under the inner Poincar\'e group.

\section{Asymptotic States: Matter Fields}
In this section we generalize canonical quantization to both a free scalar and a free Dirac field living on ${\bf M}^{\sl 4}\times {\bf M}^{\sl 4}$. Solving the field equations we express all: the fields, the inertial energy-momentum and the inner momentum constructed in the preceding section in terms of creation and annihilation operators. The field quanta corresponding to the free scalar and Dirac fields carry both inertial energy-momentum and inner momentum. We then construct the Fock spaces and the propagators belonging to both sorts of fields. Finally we introduce the concept of "gravitational limit" to connect the field quanta with observable particles.

As we aim in this paper to define a viable $S$-matrix for quantum gravity we have to first construct the Fock spaces of the asymptotic $in$- and $out$-states which are labeled by a complete set of observables such as their mass, energy-momentum, spin, charge etc. Note that these asymptotic $in$- and $out$-states exist as the gauge theory of volume-preserving diffeomorphisms coupled to all other SM fields is not asymptotically free as we have shown at one loop in \cite{chw4}. We now turn to the construction of the asymptotic $in$- and $out$-states for scalars, Dirac spinors and - in the next section - gauge vector fields respectively.

\subsection{Scalar Field}
Let us start with the Lagrangian density for a free scalar field defined on ${\bf M}^{\sl 4}\times {\bf M}^{\sl 4}$ in analogy to the Dirac case
\beq \label{17}
{\cal L}_S (\va, \pa_\mu \va) =
\frac{1}{2}\pa_\mu \va(x,X)\cdot \pa^\mu \va(x,X)
- \frac{m^2}{2} \va^2 (x,X).
\eeq
Note that by minimal coupling $\pa_\mu \ar D_\mu $ we would obtain the gauge invariant Lagrangian density including interaction terms.

The Euler-Lagrange equation for the free field is
\beq \label{18}
\left(-\pa^2 - m^2\right) \va(x,X) = 0.
\eeq

To canonically quantize we first calculate the canonically conjugate field momentum
\beq \label{19}
\Pit(x,X) = \frac{\pa{\cal L}_S}{\pa \left( \pa^{\sl 0}\va(x,X) \right)}
= \pa_{\sl 0}\va(x,X)
\eeq
and impose generalized canonical equal-time commutation relations
\beqq \label{20}
\left[ \va(t,\underline{x};X), \va(t,\underline{y};Y) \right] &=& 0
\nonumber \\
\left[ \Pit(t,\underline{x};X), \Pit(t,\underline{y};Y) \right] &=& 0 \\
\left[ \va(t,\underline{x};X), \Pit(t,\underline{y};Y) \right] &=& 
i\,\La^4\, \de^4(X-Y)\, \de^3(\underline{x}- \underline{y}). \nonumber
\eeqq
Note the appearance of factors of $\La$ to ensure the correct canonical field dimensions.

Next we "solve" the field equation Eqn.(\ref{18}) by Fourier transforming
\beq \label{21}
\va(x,X) = \intko \intK\, \La^4
\left\{ a(\underline{k},K)\, e^{-ikx - iKX} + \mbox{h.c.}
\right\}
\eeq
and keeping $k^\mu = (\om_k, \underline k)$ on the mass shell
\beq \label{22}
k^2 = m^2 \quad \mbox{or} \quad \om_k = \sqrt{{\underline k}^2 + m^2}.
\eeq

Note that only the inertial or space-time field modes are "on-shell" and obey the dispersion relation Eqn.(\ref{22}) as the field in inertial $x$-space obeys the Lagrange equation Eqn.(\ref{18}). The inner modes on the other hand do not live on a definite mass shell. This is because the inner $X$-space and the fields defined on this inner $X$-space are introduced to carry a representation of the volume-preserving diffeomorphisms of an $M^4$ in field space. Fourier-transformation of the field cannot be done onto a definite mass shell $K^2 = const$ which would be a three-dimensional subspace of $M^4$, but must be done onto a four-dimensional subspace of $M^4$.

Inverting Eqn.(\ref{21}) yields for $a$
\beq \label{23}
a(\underline{k},K) = i\, \int\! d^{\sl 3}x \intX\La^{-4}\, e^{ikx + iKX}
\, \pa_{\sl 0}\lrvec\,\, \va(t,\underline{x};X)
\eeq
with an analogous expression for $a^\dagger$. The canonical commutation relations for the $a, a^\dagger$ are easily calculated
\beqq \label{24}
\left[ a(\underline{k};K), a(\underline{h};H) \right] &=& 0
\nonumber \\
\left[ a^\dagger (\underline{k};K), a^\dagger (\underline{h};H) \right] &=& 0 \\
\left[ a (\underline{k};K), a^\dagger (\underline{h};H) \right] &=& 
2\, \om_k\, \La^{-4}\, (2\pi)^4\de^4(K-H)\,
(2\pi)^3\de^3(\underline{k}- \underline{h}). \nonumber
\eeqq
Note that without loss of generality it is possible to restrict the support of the $a, a^\dagger$ in inner momentum space 
\beq \label{25}
\mbox{supp} \Big(a (\underline{k};K)\Big) = {\bf R}^{\sl 3}\times \Big({\bf V^+}(K)\cup {\bf V^-}(K)\Big)
\eeq 
to time- and light-like vectors, where 
\beq \label{26}
{\bf V^\pm}(K) = \{K\in {\bf M^{\sl 4}}\mid K^2 \geq 0,\: \pm K^{\sl 0}
\geq 0\}.
\eeq
This is a natural restriction in view of gravitational and inertial energy-momentum being observationally equal and which for the gauge fields is a condition equivalent to its field energy being positive \cite{chw2}.

The canonical energy-momentum tensor is calculated as usual
\beqq \label{27}
\Th^{\mu\nu}(x) &=& \intX\La^{-4}\,
\Bigg\{ \pa^\mu \va(x,X)\cdot \pa^\nu \va(x,X) \\
& & -\quad \eta^{\mu\nu}
\left(\frac{1}{2}\pa_\rho \va(x,X)\cdot \pa^\rho \va(x,X)
- \frac{m^2}{2} \va^2(x,X) \right) \Bigg\} \nonumber
\eeqq
and yields the conserved inertial energy-momentum vector operator
\beq \label{28}
{\bf p}^\nu = \int\! d^{\sl 3}x \Th^{{\sl 0}\nu}(x).
\eeq
Note that letters in bold face indicate quantum operators in this paper.

Next we re-express both the momentum three vector
\beqq \label{29}
{\bf p}^i &=& \int\! d^{\sl 3}x \intX\La^{-4}\,
\pa^{\sl 0} \va(x,X)\cdot \pa^i \va(x,X) \\
&=& \frac{1}{2} \intko \intK\, \La^4\,
k^i\, \left\{ a (\underline{k};K) a^\dagger (\underline{k};K)
+ \mbox{h.c.} \right\} \nonumber
\eeqq
and the energy
\beqq \label{30}
{\bf p}^{\sl 0} &=& {\bf H} = \frac{1}{2}\, \int\! d^{\sl 3}x \intX\La^{-4}\,
\Big\{ \pa^{\sl 0} \va(x,X)\cdot \pa^{\sl 0} \va(x,X) \nonumber \\
& & +\quad \pa^i \va(x,X)\cdot \pa^i \va(x,X)
+ m^2\, \va^2(x,X) \Big\} \\
&=& \frac{1}{2} \intko \intK\, \La^4\,
\om_k\, \left\{ a (\underline{k};K) a^\dagger (\underline{k};K)
+ \mbox{h.c.} \right\} \nonumber
\eeqq
in terms of the $a, a^\dagger$. The two expressions can be combined into the covariant inertial energy-momentum vector
\beq \label{31}
{\bf p}^\nu = \frac{1}{2} \intko \intK\, \La^4\,
k^\nu\, \left\{ a (\underline{k};K) a^\dagger (\underline{k};K)
+ \mbox{h.c.} \right\}.
\eeq

In an analogous way we next determine the inner momentum tensor
\beq \label{32}
J^\mu\,_\al(x) = \intX\La^{-4}\,
\pa^\mu \va(x,X)\cdot \nabla_\al \va(x,X)
\eeq
and the conserved inner momentum vector
\beqq \label{33}
{\bf P}_\al &=& \int\! d^{\sl 3}x J^{\sl 0}\,_\al(x) \nonumber \\
&=& \int\! d^{\sl 3}x \intX\La^{-4}\,
\pa^{\sl 0} \va(x,X)\cdot \nabla_\al \va(x,X) \\
&=& \frac{1}{2} \intko \intK\, \La^4\,
K_\al \, \left\{ a (\underline{k};K) a^\dagger (\underline{k};K)
+ \mbox{h.c.} \right\}. \nonumber
\eeqq
All of the above formula are correct up to the introduction of normal ordering which we do not discuss here as no new features arise.

The calculation of the commutators of inertial energy-momentum and inner momentum with $a^\dagger$ yields
\beqq \label{34}
\left[{\bf p}_\mu, a^\dagger (\underline{k};K) \right] &=& k_\mu\, a^\dagger (\underline{k};K) \\
\left[{\bf P}_\al, a^\dagger (\underline{k};K) \right] &=& K_\al\, a^\dagger (\underline{k};K). \nonumber
\eeqq
With the help of these commutators we directly cross-check the conservation of both types of momenta
\beq \label{35}
\left[{\bf H}, {\bf p}_\mu \right] = 0, \quad \left[{\bf H}, {\bf P}_\al \right] = 0
\eeq
and derive the usual particle interpretation starting with a vacuum state
\beq \label{36}
\midd 0\rangle \quad\mbox{with}\quad \langle 0 \midd 0\rangle = 1
\eeq
which is annihilated by the destruction operator $a$
\beq \label{37}
a (\underline{k};K) \midd 0\rangle = 0
\eeq
and out of which the creation operator $a^\dagger$ generates one-particle states \beq \label{38}
a^\dagger (\underline{k};K) \midd 0\rangle = \mbox{one-particle state} 
\eeq
with definite energy-momentum $k$, inner momentum $K$ and with normalization
\beq \label{39}
\langle 0\midd a (\underline{h};H) a^\dagger (\underline{k};K) \midd 0\rangle
= 2\, \om_k\, \La^{-4}\, (2\pi)^4\de^4(K-H)\,
(2\pi)^3\de^3(\underline{k}- \underline{h}).
\eeq

Acting on a simultaneous eigenstate $\midd h,H\rangle$ of ${\bf p}_\mu$ and ${\bf P}_\al$ with eigenvalues $h_\mu$ and $H_\al$ respectively we find
\beqq \label{40}
{\bf p}_\mu\, a^\dagger (\underline{k};K) \midd h,H\rangle &=&
(h_\mu + k_\mu)\, a^\dagger (\underline{k};K) \midd h,H\rangle \nonumber \\
{\bf p}_\mu\, a(\underline{k};K) \midd h,H\rangle &=&
(h_\mu - k_\mu)\, a(\underline{k};K) \midd h,H\rangle \\
{\bf P}_\al\, a^\dagger (\underline{k};K) \midd h,H\rangle &=&
(H_\al + K_\al)\, a^\dagger (\underline{k};K) \midd h,H\rangle \nonumber \\
{\bf P}_\al\, a(\underline{k};K) \midd h,H\rangle &=&
(H_\al - K_\al)\, a(\underline{k};K) \midd h,H\rangle \nonumber
\eeqq
which allows us to construct the Fock space of stationary states by applying multiple creation operators on the vacuum state. Due to Eqn.(\ref{39}) the stationary states have positive norm as required for a consistent probabilistic interpretation of the theory.

Finally we calculate the time-ordered product of two field operators to obtain the Feynman propagator for free scalar fields
\beqq \label{41}
& & \langle 0\midd T\left(\va(x,X) \va(y,Y) \right)\midd 0\rangle
= i\, \La^4\, \de^4(X-Y)\cdot \\
& &\quad\quad\quad\quad \intk e^{-ik(x-y)}\, \frac{1}{k^2 - m^2 + i\ep} \nonumber
\eeqq
which is local in inner space.

As already pointed out we observe in Nature only particles with the same values for inertial and gravitational mass or in an arbitrary Lorentz frame with the same values for inertial and gravitational energy-momentum. This leads us to introduce the "gravitational limit" by mapping $K$ onto $k$
\beq \label{42}
\bar a^\dagger (\underline{k}) =\lim_{K \ar k} a^\dagger (\underline{k};K).
\eeq

The normalization of such states
\beq \label{43}
\left[ \bar a (\underline{k}), \bar a^\dagger (\underline{h}) \right]
=\lim_{K \ar k} \lim_{H \ar h}
2\, \om_k\, \La^{-4}\, (2\pi)^4\de^4(K-H)\, (2\pi)^3\de^3(\underline{k}- \underline{h})
\eeq
involves a regularization of the inner volume
\beq \label{44}
\de^4(0) \sim \frac{1}{(2\pi)^4}\, \intX \ar \frac{\Vreg}{(2\pi)^4}
\eeq
after which we find
\beqq \label{45}
\left[ \bar a (\underline{k}), \bar a^\dagger (\underline{h}) \right]
&=& 2\, \om_k\, \frac{\Vreg}{\La^4}\, (2\pi)^3\de^3(\underline{k}- \underline{h}) \\
&=& 2\, \om_k\, (2\pi)^3\de^3(\underline{k}- \underline{h}) \nonumber
\eeqq
using the fact that $\La$ is an a priori unspecified parameter which we can freely choose so that
\beq \label{46}
\frac{\Vreg}{\La^4} = 1
\eeq
holds. After such a choice we can still rescale $\La$ using the scaling properties of the Lagrangian and the quantum Green functions discussed in \cite{chw2,chw4}.

Hence $\bar a, \bar a^\dagger$ are properly normalized destruction and creation operators belonging to a free scalar field (but with field quanta still remembering that they carry both inertial and inner energy-momentum). These are the operators generating the field quanta which represent observable scalar particles in our approach.

\subsection{Dirac Field}
Let us start with the symmetric Lagrangian density for a free Dirac field defined on ${\bf M}^{\sl 4}\times {\bf M}^{\sl 4}$
\beqq \label{47}
{\cal L}_D (\psi, \pa_\mu \psi) &=&
\frac{i}{2}\left\{\bar\psi(x,X) (\dsl\rvec\psi(x,X))
- (\bar\psi(x,X) \dsl\lvec)\psi(x,X) \right\} \\
& &\quad - \, m\, \bar\psi(x,X) \psi(x,X). \nonumber
\eeqq
Note that by minimal coupling $\pa_\mu \ar D_\mu $ we would obtain the Lagrangian density including interaction terms.

The Euler-Lagrange equation for the free Dirac field is
\beq \label{48}
\left( i\,\dsl\rvec - m\right) \psi(x,X) = 0.
\eeq

To quantize we first "solve" the field equation Eqn.(\ref{48}) by Fourier transforming the field
\beqq \label{49}
& & \psi(x,X) = \intkm \intK\, \La^4\, \sum_{s = 1,2} \\
& &\quad \left\{ b(\underline{k},s; K) u(k,s)\, e^{-ikx - iKX} 
+ d^\dagger(\underline{k},s; K) v(k,s)\, e^{ikx + iKX} 
\right\} \nonumber
\eeqq
and putting $k^\mu$ on the mass shell
\beq \label{50}
k^2 = m^2 \quad \mbox{or} \quad k^{\sl 0} = \sqrt{{\underline k}^2 + m^2}.
\eeq
Above $s$ denotes the spin, $u(k,s)$ and $v(k,s)$ denote solutions of the free Dirac equation in $k$-space with the usual normalizations \cite{cli}.

Inversion yields for $b$ and $d^\dagger$ respectively
\beqq \label{51}
b(\underline{k},s; K) &=& \int\! d^{\sl 3}x \intX\La^{-4}\, e^{ikx + iKX}
\bar u(k,s) \ga^{\sl 0} \psi(t,\underline{x};X) \\
d^\dagger(\underline{k},s; K) &=& \int\! d^{\sl 3}x \intX\La^{-4}\,
e^{-ikx - iKX} \bar v(k,s) \ga^{\sl 0} \psi(t,\underline{x};X). \nonumber
\eeqq
We note that without loss of generality it is possible to restrict the support of the $b, d^\dagger$ and their conjugates in inner momentum space to time- and light-like vectors
\beq \label{52}
\mbox{supp} \Big( b(\underline{k},s; K) \Big) = \mbox{supp} \Big(d^\dagger (\underline{k},s; K) \Big) = {\bf R}^{\sl 3}\times \Big({\bf V^+}(K)\cup {\bf V^-}(K)\Big).
\eeq 

The canonical energy-momentum tensor is calculated as usual
\beq \label{53}
\Th^{\mu\nu}(x) = \intX\La^{-4}\,
\frac{i}{2} \bar\psi(x,X) \ga^\mu \pa^\nu\lrvec\,\, \psi(x,X)
\eeq
and yields the conserved inertial energy-momentum vector
\beqq \label{54}
& & {\bf p}^\nu = \intkm \intK\, \La^4\, k^\nu\, \sum_{s = 1,2} \\
& &\quad \left\{ b^\dagger(\underline{k},s; K) b(\underline{k},s; K)  
- d(\underline{k},s; K) d^\dagger(\underline{k},s; K \right\}. \nonumber
\eeqq

To consistently quantize we require the $b,b^\dagger$ and the $d,d^\dagger$ to fulfil the following anti-commutation relations
\beqq \label{55}
& & \left\{ b(\underline{k},s; K), b^\dagger(\underline{h},t; H) \right\} \nonumber \\
& & = \left\{ d(\underline{k},s; K), d^\dagger(\underline{h},t; H) \right\} \\
& & = \frac{k^{\sl 0}}{m}\, \de_{st}\, \La^{-4}\, (2\pi)^4\de^4(K-H)\,
(2\pi)^3\de^3(\underline{k}- \underline{h}). \nonumber
\eeqq
It is then easy to deduct the corresponding anti-commutation relation for the field components
\beq \label{56}
\left\{ \psi_\xi(t,\underline{x}; X),
\psi^\dagger_\eta(t,\underline{y}; Y) \right\}
= \de_{\xi\eta}\, \La^4\, \de^4(X-Y)\,
\de^3(\underline{x}- \underline{y}).
\eeq

We next determine the inner momentum tensor 
\beq \label{57}
J^\mu\,_\al(x) = \intX\La^{-4}\,
\frac{i}{2} \bar\psi(x,X) \ga^\mu \nabla_\al\!\lrvec\,\, \psi(x,X)
\eeq
and - up to normal ordering - the conserved inner momentum vector
\beqq \label{58}
& & {\bf P}_\al = \intkm \intK\, \La^4\, K_\al\, \sum_{s = 1,2} \\
& &\quad \left\{ b^\dagger(\underline{k},s; K) b(\underline{k},s; K)  
+ d^\dagger(\underline{k},s; K) d(\underline{k},s; K \right\}. \nonumber
\eeqq

The calculation of the commutators of inertial energy-momentum and inner momentum with $b^\dagger$ and $d^\dagger$ yields
\beqq \label{59}
\left[{\bf p}_\mu, b^\dagger (\underline{k},s; K) \right] &=& k_\mu\, b^\dagger (\underline{k},s; K) \nonumber \\
\left[{\bf p}_\mu, d^\dagger (\underline{k},s; K) \right] &=& k_\mu\, d^\dagger (\underline{k},s; K) \\
\left[{\bf P}_\al, b^\dagger (\underline{k},s; K) \right] &=& K_\al\, b^\dagger (\underline{k},s; K) \nonumber \\
\left[{\bf P}_\al, d^\dagger (\underline{k},s; K) \right] &=& K_\al\, d^\dagger (\underline{k},s; K). \nonumber
\eeqq

The usual particle interpretation is obtained starting with a vacuum state
\beq \label{60}
\midd 0\rangle \quad\mbox{with}\quad \langle 0 \midd 0\rangle = 1
\eeq
which is annihilated by the destruction operators $b,d$
\beq \label{61}
b (\underline{k},s; K) \midd 0\rangle = 0, \quad
d (\underline{k},s; K) \midd 0\rangle = 0
\eeq
and out of which the creation operators $b^\dagger,d^\dagger$ generate one-particle and -antiparticle states respectively
\beqq \label{62}
& & b^\dagger (\underline{k},s; K) \midd 0\rangle = \mbox{one-particle state} \\
& & d^\dagger (\underline{k},s; K) \midd 0\rangle = \mbox{one-antiparticle state} \nonumber
\eeqq
with definite energy-momentum $k$, inner momentum $K$, spin $s$ and with normalizations
\beqq \label{63}
& & \langle 0\midd b(\underline{k},s; K) b^\dagger(\underline{h},t; H) \midd 0\rangle \nonumber \\
& & = \langle 0\midd d(\underline{k},s; K) d^\dagger(\underline{h},t; H) \midd 0\rangle \\
& & = \frac{k^{\sl 0}}{m}\, \de_{st}\, \La^{-4}\, (2\pi)^4\de^4(K-H)\,
(2\pi)^3\de^3(\underline{k}- \underline{h}). \nonumber
\eeqq

Acting on a simultaneous eigenstate $\midd h,H\rangle$ of ${\bf p}_\mu$ and ${\bf P}_\al$ with eigenvalues $h_\mu$ and $H_\al$ respectively we find
\beqq \label{64}
{\bf p}_\mu\, b^\dagger (\underline{k},s; K) \midd h,H\rangle &=&
(h_\mu + k_\mu)\, b^\dagger (\underline{k},s; K) \midd h,H\rangle \nonumber \\
{\bf p}_\mu\, b(\underline{k},s; K) \midd h,H \rangle &=&
(h_\mu - k_\mu)\, b(\underline{k},s; K) \midd h,H\rangle \\
{\bf P}_\al\, b^\dagger (\underline{k},s; K) \midd h,H\rangle &=&
(H_\al + K_\al)\, b^\dagger (\underline{k},s; K) \midd h,H \rangle \nonumber \\
{\bf P}_\al\, b(\underline{k},s; K) \midd h,H\rangle &=&
(H_\al - K_\al)\, b(\underline{k},s; K) \midd h,H\rangle \nonumber
\eeqq
and similar relations for $d,d^\dagger$ which allows us to construct the Fock space of stationary states by applying multiple creation operators on the vacuum state. Due to Eqns.(\ref{63}) the stationary states have positive norm as required for a consistent probabilistic interpretation of the theory.

Next we calculate the time-ordered product of two field operators to obtain the Feynman propagator for free Dirac fields
\beqq \label{65}
& & \langle 0\midd T\left(\psi(x,X) \bar\psi(y,Y) \right)\midd 0\rangle
= i\, \La^4\, \de^4(X-Y)\cdot \\
& &\quad\quad\quad\quad \intk e^{-ik(x-y)}\, \frac{\ksl + m}{k^2 - m^2 + i\ep} \nonumber
\eeqq
which is local in inner space.

Finally we introduce the "gravitational limit" by mapping $K$ onto $k$
\beqq \label{66}
& & \bar b^\dagger (\underline{k},s) =\lim_{K \ar k} b^\dagger (\underline{k},s; K) \\
& & \bar d^\dagger (\underline{k},s) =\lim_{K \ar k} d^\dagger (\underline{k},s; K).  \nonumber
\eeqq

The normalization of these states involves again a regularization of the inner volume
\beqq \label{67}
& & \left[ \bar b (\underline{k},s), \bar b^\dagger (\underline{h},t) \right]
= \left[ \bar d (\underline{k},s), \bar d^\dagger (\underline{h},t) \right] \nonumber \\
& & =\lim_{K \ar k} \lim_{H \ar h}
\frac{k^{\sl 0}}{m}\, \de_{st}\, \La^{-4}\, (2\pi)^4\de^4(K-H)\,
(2\pi)^3\de^3(\underline{k}- \underline{h}) \\
& & =\frac{k^{\sl 0}}{m}\, \de_{st}\, \frac{\Vreg}{\La^4}\,
(2\pi)^3\de^3(\underline{k}- \underline{h})
=\frac{k^{\sl 0}}{m}\, \de_{st}\,
(2\pi)^3\de^3(\underline{k}- \underline{h}) \nonumber
\eeqq
and yields properly normalized destruction and creation operators belonging to a free Dirac field (but with field quanta still remembering that they carry both inertial and inner energy-momentum).

\section{Asymptotic States: Gauge Field}
In this section we quantize the free volume-preserving diffeomorphism group gauge field. Solving the linearized field equations we express all: the fields, the inertial energy-momentum and the inner momentum constructed in section 2 in terms of creation and annihilation operators and appropriate polarization vectors. The field quanta carry both inertial energy-momentum and inner momentum. We then construct the gauge field Fock space and the propagator. Finally we establish the "gravitational limit" to connect the field quanta with observable gauge particles.

Let us start with the Lagrangian density Eqn.(\ref{16}) for the gauge field defined on ${\bf M}^{\sl 4}\times {\bf M}^{\sl 4}$
\beqq \label{68}
& & {\cal L}_G (A_\mu\,^\al, \pa^\nu A_\mu\,^\al; \nabla_\be A_\mu\,^\al) = \frac{1}{4\, \La^2}\, F_{\mu\nu}\,^\al(x,X) \cdot F^{\mu\nu}\,_\al(x,X) \\
& &\quad + \frac{\la}{2\, \La^2}\, \pa_\mu A^\mu\,_\al(x,X) \cdot \pa^\nu A_\nu\,^\al(x,X) - \frac{\mu^2}{2\, \La^2}\, A^\mu\,_\al(x,X) \cdot A_\mu\,^\al(x,X) \nonumber
\eeqq
adding both a gauge-fixing and a mass term to be able to canonically quantize and to deal with eventual infrared problems \cite{cli}.

The Euler-Lagrange equation for the gauge field is
\beq \label{69}
\left( -\pa^2 - \mu^2\right) A^\mu\,_\al(x,X)
+ (1 - \la) \pa^\mu (\pa_\rho A^\rho\,_\al(x,X)) = J^\mu\,_\al(x,X),
\eeq
where both gauge field self-interactions and interactions with matter currents have been collected in the current $J_\mu\,^\al$
which by construction is divergence-free in space-time as well as in inner space
\beq \label{70}
\pa^\mu J_\mu\,^\al(x,X) = 0,\quad \nabla_\al J_\mu\,^\al(x,X) = 0.
\eeq

Taking the space-time divergence of Eqn.(\ref{69})
\beq \label{71}
\left( -\pa^2 - \frac{\mu^2}{\la}\right) (\pa_\rho A^\rho\,_\al(x,X)) = 0
\eeq
we find that with or without interactions present the divergence of the gauge field $\pa_\rho A^\rho\,_\al$ is a free scalar field with mass $\frac{\mu^2}{\la}$. This in turn allows us to define the transversal part of the gauge field as usual
\beq \label{72}
A^T_\mu\,^\al(x,X) = A_\mu\,^\al(x,X)
+ \frac{\la}{\mu^2} \pa_\mu (\pa^\rho A_\rho\,^\al(x,X)),\:
\pa^\mu A^T_\mu\,^\al(x,X) = 0.
\eeq

To keep calculations simple we set the gauge parameter
\beq \label{73}
\la = 1
\eeq
in the rest of this paper. With this choice of $\la$ and with or without interactions present the transversal part of the gauge field fulfils
\beq \label{74}
\left( -\pa^2 - \mu^2\right) A^T_\mu\,^\al(x,X) = J_\mu\,^\al(x,X).
\eeq

To canonically quantize we first calculate the canonically conjugate field momenta
\beqq \label{75}
\Pit^\nu\,_\be(x,X) &=& \frac{\pa{\cal L}_G}{\pa \left( \pa^{\sl 0}
A_\nu\,^\be(x,X) \right)} \\
&=& \La^{-2} F_{\sl 0}\,^\nu\,_\be(x,X) + \La^{-2} \eta_{\sl 0}\,^\nu \pa_\rho A^\rho\,_\be(x,X) \nonumber
\eeqq
which live in the gauge algebra like the gauge fields
\beq \label{76}
\nabla_\al A_\mu\,^\al(x,X) = 0,\quad
\nabla^\be \Pit^\nu\,_\be(x,X) = 0.
\eeq

Defining the delta function transversal in inner space
\beq \label{77}
^T\!\de_{\al\be}(X-Y) = \intK \La^4\, e^{-iK(X-Y)}
\left( \eta_{\al\be} - \frac{K_\al K_\be}{K^2}\right)
\eeq
we next impose generalized canonical equal-time commutation relations
\beqq \label{78}
\left[ A_\mu\,^\al (t,\underline{x};X), A^\nu\,_\be (t,\underline{y};Y) \right] &=& 0 \nonumber \\
\left[ \Pit_\mu\,^\al (t,\underline{x};X), \Pit^\nu\,_\be (t,\underline{y};Y) \right] &=& 0 \\
\left[A_\mu\,^\al (t,\underline{x};X), \Pit^\nu\,_\be (t,\underline{y};Y) \right] &=& 
i\,\eta_\mu\,^\nu \, ^T\!\de^\al\,_\be (X-Y)\, \de^3(\underline{x}- \underline{y}) \nonumber
\eeqq
consistent with the conditions Eqns.(\ref{76}) on the gauge fields and their conjugate momenta.

It is straightforward to re-express the canonical commutation relations Eqns.(\ref{78}) in terms of the gauge fields and their first time derivative
\beqq \label{79}
\left[ A_\mu\,^\al (t,\underline{x};X), A^\nu\,_\be (t,\underline{y};Y) \right] &=& 0 \nonumber \\
\left[ \dot A_\mu\,^\al (t,\underline{x};X), \dot A^\nu\,_\be (t,\underline{y};Y) \right] &=& 0 \\
\left[ A_\mu\,^\al (t,\underline{x};X), \dot A^\nu\,_\be (t,\underline{y};Y) \right] &=& 
i\,\eta_\mu\,^\nu \, \La^2 \,\, ^T\!\de^\al\,_\be (X-Y)\, \de^3(\underline{x}- \underline{y}) \nonumber
\eeqq
indicating that the sixteen $A_\mu\,^\al$ are quantized like a set of scalar fields. Some, however, will have negative norm in Hilbert space, a fact we will have to deal with below to arrive at a sensible theory.

Next we "solve" the field equation Eqn.(\ref{69}) and the constraint Eqn.(\ref{70}) for $\la = 1$ by Fourier transforming
\beqq \label{80}
& & A_\rho\,^\al (x,X) = \intko \intK\, \La^4\,
\sum_{\ga = 0}^3\, \sum_{\Ga = 1}^3 \\
& &\quad \ep_\rho (k,\ga)\, \Ep^\al (K,\Ga)
\left\{ a_+ (\underline{k},\ga;K,\Ga)\, e^{-ikx - iKX} + \mbox{h.c.} \right\}
\nonumber
\eeqq
and putting $k^\mu$ on the mass shell
\beq \label{81}
k^2 = \mu^2 \quad \mbox{or} \quad \om_k = \sqrt{{\underline k}^2 + \mu^2}.
\eeq
Above $\ga = 0,1,2,3$ and $\Ga = 1,2,3$ denote the polarization degrees of freedom,
\beq \label{82}
\ep_\rho (k,0) = \frac{k_\rho}{\mu}\quad \mbox{and}\quad \ep_\rho (k,\ga),\quad \ga = 1,2,3
\eeq
and
\beq \label{83}
\Ep^\al (K,\Ga),\quad \Ga = 1,2,3
\eeq
the corresponding polarization vectors in space-time and inner space. Note that $\Ga = 0$ does not appear due to the transversality $\nabla_\al A_\mu\,^\al(x,X) = 0$ of the gauge field in inner space.

$\ep_\rho$ and $\Ep^\al$ obey the following relations
\beqq \label{84}
& & k^\rho\cdot \ep_\rho (k,\ga \neq 0) = 0, \quad
\ep_\rho (k,\ga)\cdot \ep^\rho (k,\ga') = \eta^{\ga\ga'} \\
& & \sum_{\ga = 1}^3\, \ep^\rho (k,\ga)\, \ep^\si (k,\ga) =  -\eta^{\rho\si} + \frac{k^\rho k^\si}{\mu^2} \nonumber
\eeqq
and
\beqq \label{85}
& & K_\al\cdot \Ep^\al (K,\Ga) = 0, \quad
\Ep^\al (K,\Ga)\cdot \Ep_\al (K,\Ga') = \eta^{\Ga\Ga'} \\
& & \sum_{\Ga = 1}^3\, \Ep_\al (K,\Ga)\, \Ep_\be (K,\Ga) =  -\eta_{\al\be} + \frac{K_\al K_\be}{K^2}\quad \mbox{for}\quad K^2 \geq 0. \nonumber
\eeqq

To ensure positivity of the gauge field energy the support of the $a_+$ is restricted to time- and light-like vectors in inner space \cite{chw2}
\beq \label{86}
\mbox{supp} \Big(a_+ (\underline{k},\ga;K,\Ga)\Big) = {\bf R}^{\sl 3}\times \Big({\bf V^+}(K)\cup {\bf V^-}(K)\Big).
\eeq 

Inversion yields for $a_+$
\beqq \label{87}
& & a_+ (\underline{k},\ga;K,\Ga) = i\, \eta^{\ga\ga}\, \eta^{\Ga\Ga}
\int\! d^{\sl 3}x \intX\La^{-4} \\
& &\quad e^{ikx + iKX}\, 
\pa_{\sl 0}\lrvec\,\, \Big( \ep^\rho (k,\ga) \Ep_\al (K,\Ga)
A_\rho\,^\al (x,X) (t,\underline{x};X) \Big) \nonumber
\eeqq 
with an analogous expression for $a_+ ^\dagger$.

The commutation relations for the $a_+, a_+ ^\dagger$ are then easily obtained
\beqq \label{88}
& & \Big[ a_+ (\underline{k},\ga;K,\Ga), a_+ (\underline{h},\ga';H,\Ga') \Big] = 0 \nonumber \\
& & \Big[ a_+ ^\dagger(\underline{k},\ga;K,\Ga),
a_+ ^\dagger(\underline{h},\ga';H,\Ga') \Big] = 0 \\
& & \Big[ a_+ (\underline{k},\ga;K,\Ga), a_+ ^\dagger(\underline{h},\ga';H,\Ga') \Big] = 2\, \om_k\, \eta^{\ga\ga'}\, \eta^{\Ga\Ga'} \nonumber \\
& &\quad \La^{-2}\, (2\pi)^4\de^4(K-H)\,
(2\pi)^3\de^3(\underline{k}- \underline{h}) \nonumber
\eeqq
and show that states have positive norm only for $\ga, \ga' = 1,2,3$.

The canonical energy-momentum tensor is calculated as usual
\beqq \label{89}
\Th^{\mu\nu}(x) &=& \intX\La^{-6}\, \Bigg\{
F^{\mu\rho}\,_\al(x,X) \cdot \pa^\nu A_\rho\,^\al(x,X) \nonumber \\
& &\quad - \frac{1}{4}\, \eta^{\mu\nu} F_{\rho\si}\,^\al(x,X) \cdot F^{\rho\si}\,_\al(x,X) \nonumber \\
& &\quad + \pa_\rho A^\rho\,_\al(x,X) \cdot \pa^\nu A^\mu\,^\al(x,X) \\
& &\quad - \frac{1}{2}\, \eta^{\mu\nu} \pa_\rho A^\rho\,_\al(x,X) \cdot \pa^\si A_\si\,^\al(x,X) \nonumber \\
& &\quad - \frac{\mu^2}{2}\, \eta^{\mu\nu} A^\rho\,_\al(x,X) \cdot A_\rho\,^\al(x,X)
\Bigg\}. \nonumber
\eeqq

Next we re-express both the momentum three vector
\beq \label{90}
{\bf p}^i = \int\! d^{\sl 3}x \intX\La^{-6}\,
\pa^{\sl 0} A^\rho\,_\al(x,X) \cdot \pa^i A_\rho\,^\al(x,X) \\
\eeq
and the energy
\beqq \label{91}
{\bf p}^{\sl 0} &=& \frac{1}{2}\, \int\! d^{\sl 3}x \intX\La^{-4}\,
\Big\{ \pa^{\sl 0} A^\rho\,_\al(x,X) \cdot \pa^{\sl 0} A_\rho\,^\al(x,X) \\
& & \quad + \pa^i A^\rho\,_\al(x,X) \cdot \pa^i A_\rho\,^\al(x,X) 
+ \mu^2\, A^\rho\,_\al(x,X) \cdot A_\rho\,^\al(x,X) \Big\} \nonumber
\eeqq
in terms of the $a_+, a_+ ^\dagger$ obtaining the covariant inertial energy-momentum vector
\beqq \label{92}
{\bf p}^\nu &=& \frac{1}{2} \intko \intK\, \La^2\,
k^\nu\, \sum_{\ga = 0}^3\, \sum_{\Ga = 1}^3 \\
& &\quad \eta^{\ga\ga}\, \eta^{\Ga\Ga}
\left\{ a_+ (\underline{k},\ga;K,\Ga) a_+ ^\dagger(\underline{k},\ga;K,\Ga)
+ \mbox{h.c.} \right\}. \nonumber
\eeqq
Note that the energy ${\bf p}^{\sl 0} > 0$ is positive if the summation is extended over $\ga = 1,2,3$ only which is the case for purely transversal states.

In an analogous way we next determine the inner momentum tensor
\beqq \label{93}
J^\mu\,_\be(x) &=& \intX\La^{-6}\,
\Big\{ F^{\mu\rho}\,_\al(x,X) \cdot \nabla_\be A_\rho\,^\al(x,X) \\
& &\quad + \pa_\rho A^\rho\,_\al(x,X) \cdot \nabla_\be A^\mu\,^\al(x,X) \Big\} \nonumber
\eeqq
and the conserved inner momentum vector
\beqq \label{94}
{\bf P}_\be &=& \int\! d^{\sl 3}x \intX\La^{-6}\,
\pa^{\sl 0} A^\rho\,_\al(x,X) \cdot \nabla_\be A_\rho\,^\al(x,X) \nonumber \\
&=& \frac{1}{2} \intko \intK\, \La^2\,
K_\be\, \sum_{\ga = 0}^3\, \sum_{\Ga = 1}^3 \\
& &\quad \eta^{\ga\ga}\, \eta^{\Ga\Ga}
\left\{ a_+ (\underline{k},\ga;K,\Ga) a_+ ^\dagger(\underline{k},\ga;K,\Ga)
+ \mbox{h.c.} \right\}. \nonumber
\eeqq
All of the above formula are correct again up to the introduction of normal ordering.

The calculation of the commutators of the inertial energy-momentum and the inner momentum with $a_+ ^\dagger$ yields
\beqq \label{95}
\left[{\bf p}_\mu, a_+ ^\dagger(\underline{k},\ga;K,\Ga) \right] &=& \eta^{\ga\ga}\, \eta^{\Ga\Ga}\, k_\mu\, a_+ ^\dagger(\underline{k},\ga;K,\Ga) \\
\left[{\bf P}_\al, a_+ ^\dagger(\underline{k},\ga;K,\Ga)\right] &=& \eta^{\ga\ga}\, \eta^{\Ga\Ga}\, K_\al\, a_+ ^\dagger(\underline{k},\ga;K,\Ga) \nonumber
\eeqq
allowing for an easy cross-check of the conservation of both types of momenta
\beq \label{96}
\left[{\bf H}, {\bf p}_\mu \right] = 0, \quad \left[{\bf H}, {\bf P}_\al \right] = 0.
\eeq

We establish the usual particle interpretation starting with a vacuum state
\beq \label{97}
\midd 0\rangle \quad\mbox{with}\quad \langle 0 \midd 0\rangle = 1
\eeq
which is annihilated by the destruction operator $a_+$
\beq \label{98}
a_+ (\underline{k},\ga;K,\Ga) \midd 0\rangle = 0
\eeq
and out of which the creation operator $a_+ ^\dagger$ generates one-particle states 
\beq \label{99}
a_+ ^\dagger(\underline{k},\ga;K,\Ga) \midd 0\rangle = \mbox{one-particle state} \nonumber
\eeq
with definite energy-momentum $k$, inner momentum $K$ and polarizations $\ga$, $\Ga$ and with normalization
\beqq \label{100}
& & \langle 0\midd a_+ (\underline{h},\ga';H,\Ga') 
a_+ ^\dagger(\underline{k},\ga;K,\Ga) \midd 0\rangle
= 2\, \om_k\, \eta^{\ga\ga'}\, \eta^{\Ga\Ga'} \\
& &\quad \La^{-2}\, (2\pi)^4\de^4(K-H)\,
(2\pi)^3\de^3(\underline{k}- \underline{h}). \nonumber 
\eeqq 

Acting on a simultaneous eigenstate $\midd h,H\rangle$ of ${\bf p}_\mu$ and ${\bf P}_\al$ with eigenvalues $h_\mu$ and $H_\al$ respectively we find
\beqq \label{101}
& & \!\!\!\!\!\!\!\!\!\!\!\!\!\!
{\bf p}_\mu\, a_+ ^\dagger(\underline{k},\ga;K,\Ga) \midd h,H\rangle =
\eta^{\ga\ga}\, \eta^{\Ga\Ga}\, (h_\mu + k_\mu)\, a_+ ^\dagger(\underline{k},\ga;K,\Ga) \midd h,H\rangle \nonumber \\
& & \!\!\!\!\!\!\!\!\!\!\!\!\!\!
{\bf p}_\mu\, a_+ (\underline{k},\ga;K,\Ga) \midd h,H\rangle =
\eta^{\ga\ga}\, \eta^{\Ga\Ga}\, (h_\mu - k_\mu)\, a_+ (\underline{k},\ga;K,\Ga) \midd h,H\rangle \\
& & \!\!\!\!\!\!\!\!\!\!\!\!\!\!
{\bf P}_\al\, a_+ ^\dagger(\underline{k},\ga;K,\Ga) \midd h,H\rangle =
\eta^{\ga\ga}\, \eta^{\Ga\Ga}\, (H_\al + K_\al)\, a_+ ^\dagger(\underline{k},\ga;K,\Ga) \midd h,H\rangle \nonumber \\
& & \!\!\!\!\!\!\!\!\!\!\!\!\!\!
{\bf P}_\al\, a_+ (\underline{k},\ga;K,\Ga) \midd h,H\rangle =
\eta^{\ga\ga}\, \eta^{\Ga\Ga}\, (H_\al - K_\al)\, a_+ (\underline{k},\ga;K,\Ga) \midd h,H\rangle \nonumber
\eeqq
which allows us to construct the Fock space of stationary states by applying multiple creation operators on the vacuum state, however, at the price of having to deal with an indefinite metric at this point as is usual in the Stueckelberg-Gupta-Bleuler approach for quantizing gauge fields.

Finally we calculate the time-ordered product of two field operators to obtain the Feynman propagator for free gauge fields
\beqq \label{102}
& & \langle 0\midd T\left( A_\mu\,^\al(x,X) A_\nu\,^\be(x,X)
\right)\midd 0\rangle = -i\, \La^2\,\, ^T\!\de^{\al\be}(X-Y) \cdot \\
& &\quad\quad\quad\quad \intk e^{-ik(x-y)}\, \frac{\eta_{\mu\nu}}{k^2 - \mu^2 + i\ep} \nonumber
\eeqq
which is local in inner space.

Before turning to discuss the "gravitational limit" we define the Fock sub-space of positive norm states requiring their annihilation by
$\pa^\rho A_\rho^{(-)}\,^\al$
\beq \label{103}
{\cal F}_P = \left\{ \midd \al \rangle \midd
\pa^\rho A_\rho^{(-)}\,^\al(x,X) \midd \al \rangle = 0 \right\},
\eeq
where $A_\rho^{(-)}\,^\al$ is given by
\beqq \label{104}
& & A_\rho^{(-)}\,^\al(x,X) = \intko \intK\, \La^4\,
\sum_{\ga = 0}^3\, \sum_{\Ga = 1}^3 \\
& &\quad \ep_\rho (k,\ga)\, \Ep^\al (K,\Ga)\,
a_+ (\underline{k},\ga;K,\Ga)\, e^{-ikx - iKX}
\nonumber
\eeqq
and its divergence by
\beqq \label{105}
& & \pa^\rho A_\rho^{(-)}\,^\al(x,X) = \intko \intK\, \La^4\,
\sum_{\Ga = 0}^3\, \mu \\
& &\quad \Ep^\al (K,\Ga)\, a_+ (\underline{k},0;K,\Ga)\, e^{-ikx - iKX}
\nonumber
\eeqq
containing only the destruction operators $a_+$ for $\ga = 0$. Due to Eqn.(\ref{100}) ${\cal F}_P$ contains positive norm states only 
\beq \label{106}
a_+ ^\dagger(\underline{k},\ga;K,\Ga) \midd 0\rangle = \mbox{positive norm states for}\,\, \ga,\, \Ga = 1,2,3
\eeq
as required for a consistent probabilistic interpretation of the theory.

Finally we introduce the "gravitational limit" by mapping $K$ onto $k$
\beq \label{107}
\bar a^\dagger (\underline{k},\ga,\Ga) = \lim_{K \ar k}
a_+ ^\dagger(\underline{k},\ga;K,\Ga).
\eeq

The normalization of these states involves again a regularization of the inner volume
\beqq \label{108}
& & \left[ \bar a (\underline{k},\ga,\Ga),
\bar a^\dagger (\underline{h},\ga',\Ga') \right]
=\lim_{K \ar k} \lim_{H \ar h}
2\, \om_k\, \eta^{\ga\ga'}\, \eta^{\Ga\Ga'} \nonumber \\
& &\quad\quad\quad \La^{-2}\, (2\pi)^4\de^4(K-H)\, (2\pi)^3\de^3(\underline{k}- \underline{h}) \\
& &\quad\quad = 2\, \om_k\, \eta^{\ga\ga'}\, \eta^{\Ga\Ga'}\,  \frac{\Vreg}{\La^4}\, \La^2\, (2\pi)^3\de^3(\underline{k}- \underline{h}) \nonumber \\
& &\quad\quad = 2\, \om_k\, \eta^{\ga\ga'}\, \eta^{\Ga\Ga'}\, \La^2\, (2\pi)^3\de^3(\underline{k}- \underline{h}) \nonumber
\eeqq
and yields properly normalized destruction and creation operators belonging to a free gauge field (but with field quanta still remembering that they carry both inertial and inner energy-momentum).

The physical Fock space ${\cal F}_{\ss \mbox{phys}}$
of positive norm states is again obtained requiring their annihilation by $\pa^\rho \bar A_\rho^{(-)}\,^\al$
\beq \label{109}
{\cal F}_{\ss \mbox{phys}} = \left\{ \midd \be \rangle \midd
\pa^\rho \bar A_\rho^{(-)}\,^\al(x) \midd \be \rangle = 0 \right\},
\eeq
where
\beq \label{110}
\bar A_\rho^{(-)}\,^\al(x) = \intko
\sum_{\ga = 0}^3\, \sum_{\Ga = 1}^3\, \ep_\rho (k,\ga)\, \Ep^\al (k,\Ga)\,
\bar a (\underline{k},\ga,\Ga)\, e^{-ikx}
\eeq
and
\beq \label{111}
\pa^\rho \bar A_\rho^{(-)}\,^\al(x) = \intko\,
\mu \sum_{\Ga = 1}^3\, \Ep^\al (k,\Ga)\,
\bar a (\underline{k},0,\Ga)\, e^{-ikx}.
\eeq
The positive norm one-particle states
\beq \label{112}
\bar a^\dagger (\underline{k},\ga,\Ga) \midd 0\rangle = \mbox{physical states for}\,\, \ga,\, \Ga = 1,2,3
\eeq
represent then the asymptotic field quanta which correspond to the observable gravitons in our approach.

\section{The Definition of the $S$-Matrix in Quantum Gravity}
In this section we first introduce the operator ${\bf S}$ mapping the asymptotic $in$-Fock space onto the asymptotic $out$-Fock space and discuss its properties. We then define the $S$-matrix $S$ in quantum gravity as the gravitational limit of the operator ${\bf S}$ mapping the physical $in$-Fock space onto the physical $out$-Fock space. $S$ is shown to be a unitary operator on the asymptotic physical Fock space if ${\bf S}$ is on the asymptotic Fock space.

Let us define the operator ${\bf S}$ by its matrix elements
\beq \label{113}
{\bf S}_{h_j,H_j;k_i,K_i} \equiv 
\langle h_1, H_1;\dots ;h_m, H_m\, \mbox{out}\midd k_1, K_1;\dots ;k_n, K_n\, \mbox{in} \rangle
\eeq
or
\beq \label{114}
{\bf S}_{h_j,H_j;k_i,K_i} =
\langle h_1, H_1;\dots ;h_m, H_m\, \mbox{in}\midd {\bf S} \midd
k_1, K_1;\dots ;k_n, K_n\, \mbox{in} \rangle.
\eeq
Note that we subsume all quantum numbers other than $k$ and $K$ characterizing an asymptotic stationary state in the above notation as they add no new features in our context.

${\bf S}$ maps the asymptotic $in$-Fock space ${\cal F}_{\mbox{\ss in}}$ onto the asymptotic $out$-Fock space ${\cal F}_{\mbox{\ss out}}$
\beq \label{115}
\langle h_1, H_1;\dots ;h_m, H_m\, \mbox{in}\midd {\bf S}
 = \langle h_1, H_1;\dots ;h_m, H_m\, \mbox{out}\midd
\eeq
and is invertible by construction
\beq \label{116}
\langle h_1, H_1;\dots ;h_m, H_m\, \mbox{out}\midd {\bf S}^{-1}
 = \langle h_1, H_1;\dots ;h_m, H_m\, \mbox{in}\midd.
\eeq

Next we list a few of the key properties of ${\bf S}$ \cite{cli,bjd}:

\noindent A) Invariance of the vacuum yields
\beq \label{117}
\langle 0\, \mbox{in}\midd {\bf S}
 = \langle 0\, \mbox{out} \!\mid\, = e^{i\al} \langle 0\, \mbox{in} \!\mid 
\eeq
and its uniqueness allows us to set the phase equal to zero so that
\beq \label{118}
\langle 0\, \mbox{out} \!\mid\, = \langle 0\, \mbox{in} \!\mid\,
= \langle 0\, \!\mid.
\eeq

\noindent B) Invariance of the one-particle states yields
\beqq \label{119}
\langle p, P\, \mbox{in}\midd {\bf S} \midd p, P\, \mbox{in} \rangle
&=& \langle p, P\, \mbox{out}\midd p, P\, \mbox{in} \rangle \\
&=& \langle p, P\, \mbox{in} \midd p, P\, \mbox{in} \rangle = 1 \nonumber 
\eeqq
as we have
\beq \label{120}
\midd p, P\, \mbox{in} \rangle = \,\mid\! p, P\, \mbox{out} \rangle
= \,\mid\! p, P\, \rangle.
\eeq

\noindent C) The operator adjoint to ${\bf S}$ is defined by
\beq \label{121}
{\bf S}^\dagger \midd k_1, K_1;\dots ;k_n, K_n\, \mbox{in} \rangle
= \,\mid\! k_1, K_1;\dots ;k_n, K_n\, \mbox{out} \rangle
\eeq
and ${\bf S}$ is by construction unitary
\beqq \label{122}
& & \langle h_1, H_1;\dots ;h_m, H_m\, \mbox{in}\midd {\bf S}\,
{\bf S}^\dagger \midd k_1, K_1;\dots ;k_n, K_n\, \mbox{in} \rangle
\nonumber \\
& &\quad = \langle h_1, H_1;\dots ;h_m, H_m\, \mbox{out}\midd 
k_1, K_1;\dots ;k_n, K_n\, \mbox{out} \rangle \\
& &\quad = \de_{h_j,H_j;k_i,K_i} \nonumber
\eeqq
or
\beq \label{123}
{\bf S}\, {\bf S}^\dagger = {\bf S}^\dagger {\bf S} = {\bf 1}
\eeq
if the sets of ortho-normal vectors spanning the asymptotic $in$- and $out$-Fock spaces 
\beq \label{124}
\midd k_1, K_1;\dots ;k_n, K_n\, \mbox{in} \rangle ,\quad
\midd k_1, K_1;\dots ;k_n, K_n\, \mbox{out} \rangle
\eeq
in Eqn.(\ref{122}) are complete.

Note that in the case of gauge fields ${\bf S}$ is pseudo-unitary only and its unitarity on the asymptotic $in$- and $out$-Fock subspaces of positive norm states has to be proven. In the case of the gauge field $A_\mu\,^\al(x,X)$ we have established that proof in \cite{chw4}.

Next we define the $S$-matrix $S$ in quantum gravity by the gravitational limit of the matrix elements of the operator ${\bf S}$
\beqq \label{125}
& & S_{h_j,k_i} \equiv \lim_{H_j \ar h_j} \lim_{K_i \ar k_i}
{\bf S}_{h_j,H_j;k_i,K_i} \\
& &= \lim_{H_j \ar h_j} \lim_{K_i \ar k_i} \langle h_1, H_1;\dots ;h_m, H_m\, \mbox{out}\midd k_1, K_1;\dots ;k_n, K_n\, \mbox{in} \rangle \nonumber \\
& &= \langle h_1\dots h_m\, \mbox{out} \midd k_1 \dots k_n\, \mbox{in} \rangle \nonumber \\
& &= \langle h_1\dots h_m\, \mbox{in} \midd S\midd 
k_1\dots k_n\, \mbox{in} \rangle. \nonumber
\eeqq
The operator $S$ maps the asymptotic physical $in$-Fock space onto the asymptotic physical $out$-Fock space 
\beq \label{126}
\langle h_1\dots h_m\, \mbox{in}\midd S
 = \langle h_1\dots h_m\, \mbox{out}\midd
\eeq
and is invertible
\beq \label{127}
\langle h_1\dots h_m\, \mbox{out}\midd S^{-1}
 = \langle h_1\dots h_m\, \mbox{in}\midd.
\eeq

The properties A), B) and C) hold again:

\noindent A') Eqn.(\ref{117}) yields
\beq \label{128}
\langle 0\, \mbox{in}\midd S
 = \langle 0\, \mbox{out} \!\mid\, = e^{i\be} \langle 0\, \mbox{in} \!\mid
\eeq
and we can again set the phase equal to zero so that
\beq \label{129}
\langle 0\, \mbox{out} \!\mid\, = \langle 0\, \mbox{in} \!\mid\,
= \langle 0\, \!\mid. \nonumber
\eeq
\noindent B') Eqn.(\ref{119}) yields
\beqq \label{130}
\langle p\, \mbox{in}\midd S \midd p\, \mbox{in} \rangle
&=& \langle p\, \mbox{out}\midd p\, \mbox{in} \rangle \\
&=& \langle p\, \mbox{in} \midd p\, \mbox{in} \rangle = 1 \nonumber
\eeqq
as we have again
\beq \label{131}
\midd p\, \mbox{in} \rangle = \,\mid\! p\, \mbox{out} \rangle
= \,\mid\! p\, \rangle.
\eeq

\noindent C') To prove the unitarity of $S$ on the asymptotic physical Fock space we introduce the self-adjoint projection operator ${\bf P} = {\bf P}^2$ which maps any $in$- or $out$-state in the asymptotic positive-norm Fock space on its gravitational-limit state vector in the asymptotic physical Fock space 
${\cal F}_{\ss \mbox{phys}}$
\beq \label{132}
{\bf P} \midd k_1, K_1;\dots ;k_n, K_n\, \mbox{in} \rangle =
\lim_{K_i \ar k_i} \midd k_1, K_1;\dots ;k_n, K_n\, \mbox{in} \rangle.
\eeq
Note that ${\bf P}$ is the identity in the asymptotic physical Fock space
\beq \label{133}
{\bf P} = {\bf 1}_{\ss \mbox{phys}}.
\eeq

We can now express $S$ in terms of ${\bf S}$ and ${\bf P}$ as
\beq \label{134}
\langle \dots \mbox{in} \midd {\bf P} S = \langle \dots \mbox{in} \midd 
{\bf S} {\bf P}.
\eeq 
Assuming ${\bf S}$ to be unitary allows us to write
\beqq \label{135}
{\bf 1}_{\ss \mbox{phys}} &=& {\bf P}\, {\bf 1}\, {\bf P} =
{\bf P} {\bf S}^\dagger\, {\bf S} {\bf P} \\
&=& S^\dagger {\bf P}\, {\bf P}\, S = S^\dagger S. \nonumber
\eeqq
This proves the unitarity of the gravitational $S$-matrix $S$ if ${\bf S}$ is unitary which is a key pre-requisite for a viable physical theory.

\section{Generalized LSZ Reduction Formulae: Matter Fields}
In this section we generalize the LSZ reduction formulae to quantum gravity thereby expressing $S$-matrix elements as the gravitational limit of truncated Fourier-transformed vacuum expectation values of time-ordered products of field operators of the interacting theory on the mass shell in the usual way \cite{cli,bjd}. Noting that the generating functional of the latter in the presence of gauge fields has been defined and renormalized to one-loop in \cite{chw4} this puts us in a position to perturbatively calculate any $S$-matrix element containing external "matter" states.

\subsection{Scalar Field}
Following the standard approach \cite{bjd} let us start with the field equation for the interacting field
\beq \label{136}
\left(-\pa^2 - m^2\right) \va(x,X) = j(x,X).
\eeq
Here $j$ contains all the interaction terms including possible self-interactions.

We next define the $in$- and $out$-fields $\va_{\mbox{\ss in/out}}$
\beqq \label{137}
& & {\sqrt Z}\, \va_{\mbox{\ss in/out}}(x,X) = \va(x,X) \\
& &\quad - \inty\intY \La^{-4}\, \Del_{\mbox{\ss ret/adv}} (x-y, X-Y)\, j(y,Y), \nonumber
\eeqq
where $\Del_{\mbox{\ss ret/adv}}$ are the retarded and advanced Green functions
\beq \label{138}
\left(-\pa^2 - m^2\right) \Del_{\mbox{\ss ret/adv}} (x-y, X-Y)
= \La^4\, \de^4(X-Y)\, \de^4(x-y)
\eeq 
vanishing for $x^{\sl 0} < y^{\sl 0}$ and $x^{\sl 0} > y^{\sl 0}$ respectively.

By construction $\va_{\mbox{\ss in/out}}$ fulfil the Euler-Lagrange equation Eqn.(\ref{18}) for a free scalar field
\beq \label{139}
\left(-\pa^2 - m^2\right) {\sqrt Z}\, \va_{\mbox{\ss in/out}}(x,X) = 0
\eeq
which we have quantized in section 3 including the construction of the asymptotic physical Fock space of stationary states.

We leave aside the more technical discussion of the Kallen-Lehmann spectral representation for the commutator of two interacting fields which restricts $Z$ to
\beq \label{140}
0\leq Z < 1
\eeq 
as no new mathematical features arise.

Noting that $\va(x,X) \ar {\sqrt Z}\, \va_{\mbox{\ss in/out}}(x,X)$ for $x^{\sl 0}\ar \pm \infty$ can hold as usual in the mean only
\beq \label{141}
\lim_{x^{\sl 0} \ar \pm\infty} \langle \be \midd
\va (x^{\sl 0},\underline{x};X) \midd \al \rangle
= {\sqrt Z}\, \langle \be \midd \va_{\mbox{\ss out/in}} (x^{\sl 0},\underline{x};X) \midd \al \rangle
\eeq
we turn to link matrix elements of the gravitational $S$-operator to vacuum expectation values of time-ordered products of interacting field operators. 

Starting with a general $S$-matrix element $\langle \be\, \mbox{out}\midd k\, \al\, \mbox{in} \rangle$, where the $in$-state contains a one-particle state with momentum $k$, we perform the first reduction in quite some detail
\beqq \label{142}
S_{\be,k \al} &=& \langle \be\, \mbox{out}\midd k\, \al\, \mbox{in} \rangle
= \langle \be\, \mbox{out}\midd \bar a^\dagger_{\mbox{\ss in}} (\underline{k}) \midd \al\, \mbox{in} \rangle \nonumber \\
&=& \lim_{K \ar k} \langle \be\, \mbox{out}\midd
a^\dagger_{\mbox{\ss in}} (\underline{k};K) \midd \al\, \mbox{in} \rangle \nonumber \\
&=& \langle \be\, -\!k\, \mbox{out}\midd \al\, \mbox{in} \rangle \nonumber \\
& &+ \lim_{K \ar k} \langle \be\, \mbox{out}\midd
a^\dagger_{\mbox{\ss in}} (\underline{k};K)
- a^\dagger_{\mbox{\ss out}} (\underline{k};K) \midd \al\, \mbox{in} \rangle \nonumber \\
&=& \mbox{elastic contribution} \nonumber \\
& &- \lim_{K \ar k}\, i\,
\int\! d^{\sl 3}x \intX\La^{-4}\, e^{-ikx - iKX} \nonumber \\
& &\quad\quad\quad \pa_{\sl 0}\lrvec\,\, \langle \be\, \mbox{out}\midd
\va_{\mbox{\ss in}} (x^{\sl 0},\underline{x};X) -\va_{\mbox{\ss out}}
(x^{\sl 0},\underline{x};X) \midd \al\, \mbox{in} \rangle \nonumber \\
&=& \mbox{elastic contribution} \nonumber \\
& &+ \lim_{K \ar k}\, \frac{i}{\sqrt Z}\,
\Big(\lim_{x^{\sl 0} \ar \infty} - \lim_{x^{\sl 0} \ar -\infty}\Big)
\int\! d^{\sl 3}x \intX\La^{-4} \nonumber \\
& &\quad\quad\quad e^{-ikx - iKX}\, \pa_{\sl 0}\lrvec\,\, \langle \be\, \mbox{out}\midd
\va (x^{\sl 0},\underline{x};X) \midd \al\, \mbox{in} \rangle \\
&=& \mbox{elastic contribution} \nonumber \\
& &+ \lim_{K \ar k}\, \frac{i}{\sqrt Z}\,
\intx\, \frac{\pa}{\pa x^{\sl 0}} \intX\La^{-4} \nonumber \\
& &\quad\quad\quad e^{-ikx - iKX}\, \pa_{\sl 0}\lrvec\,\, \langle \be\, \mbox{out}\midd
\va (x^{\sl 0},\underline{x};X) \midd \al\, \mbox{in} \rangle \nonumber \\
&=& \mbox{elastic contribution} \nonumber \\
& &+ \lim_{K \ar k}\, \frac{i}{\sqrt Z}\,
\intx\, \intX\La^{-4}\, e^{-ikx - iKX} \nonumber \\
& &\quad\quad\quad 
\Big( {\pa\rvec}_{\sl 0}^2 - {\pa\lvec}_{\sl 0}^2 \Big)
\langle \be\, \mbox{out}\midd
\va (x^{\sl 0},\underline{x};X) \midd \al\, \mbox{in} \rangle \nonumber \\
&=& \mbox{elastic contribution} \nonumber \\
& &+ \lim_{K \ar k}\, \frac{i}{\sqrt Z}\,
\intx\, \intX\La^{-4}\, e^{-ikx - iKX} \nonumber \\
& &\quad\quad\quad 
\Big( {\pa\rvec}^2 + m^2 \Big)
\langle \be\, \mbox{out}\midd
\va (x,X) \midd \al\, \mbox{in} \rangle, \nonumber
\eeqq
where we have made crucial use of the asymptotic condition Eqn.(\ref{141}) and the on-shell condition $k^2 = m^2$.

Iterating the above steps we obtain the generalized LSZ reduction formula for scalar matter
\beqq \label{143}
& & \langle h_1\dots h_m\, \mbox{out}\midd k_1\dots k_n\, \mbox{in} \rangle
= \left( \frac{i}{\sqrt Z} \right)^{m+n} \Pi_{i = 1}^n \lim_{K_i \ar k_i} \Pi_{j = 1}^m \lim_{H_j \ar h_j} \nonumber \\
& &\quad\quad \Pi_{i = 1}^n \intx_i \intX\!_i\, \La^{-4 n}\,
\Pi_{j = 1}^m \inty_j \intY\!\!_j\, \La^{-4 m} \nonumber \\
& &\quad\quad\quad\quad\quad\quad \Pi_{i = 1}^n e^{-ik_i x_i - iK_i X_i} \Big( {\pa\rvec}_i^2 + m^2 \Big) \\
& &\quad\quad \langle 0 \midd
T\Big( \va (x_1,X_1)\dots \va (x_n,X_n) \va (y_1,Y_1)\dots \va (y_m,Y_m) \Big) \midd 0 \rangle \nonumber \\
& &\quad\quad\quad\quad\quad\quad \Pi_{j = 1}^m \Big( {\pa\lvec}_j^2 + m^2 \Big) e^{ih_j y_j + iH_j Y_j} \nonumber
\eeqq
linking general matrix elements of the gravitational $S$-operator to vacuum expectation values of time-ordered products of interacting field operators.

\subsection{Dirac Field}
Let us start with the field equation for the interacting Dirac field
\beq \label{144}
\left(i \dsl - m\right) \psi(x,X) = j(x,X)
\eeq
where $j$ contains all the interaction terms.

We next define the $in$- and $out$-fields $\psi_{\mbox{\ss in/out}}$
\beqq \label{145}
& & {\sqrt Z_2}\, \psi_{\mbox{\ss in/out}}(x,X) = \psi(x,X) \\
& &\quad - \inty\intY \La^{-4}\, S_{\mbox{\ss ret/adv}} (x-y, X-Y)\, j(y,Y). \nonumber
\eeqq
Above $S_{\mbox{\ss ret/adv}}$ are the retarded and advanced Green functions of the Dirac operator 
\beq \label{146}
\left(i \dsl - m\right) S_{\mbox{\ss ret/adv}} (x-y, X-Y)
= \La^4\, \de^4(X-Y)\, \de^4(x-y)
\eeq 
vanishing for $x^{\sl 0} < y^{\sl 0}$ and $x^{\sl 0} > y^{\sl 0}$ respectively.

By construction $\psi_{\mbox{\ss in/out}}$ fulfil the free Dirac equation Eqn.(\ref{48})
\beq \label{147}
\left(i \dsl - m\right) {\sqrt Z_2}\, \psi_{\mbox{\ss in/out}}(x,X) = 0
\eeq
which we have quantized in section 3 including the construction of the asymptotic physical Fock space of stationary states.

Noting that 
\beq \label{148}
\lim_{x^{\sl 0} \ar \pm\infty} \langle \be \midd
\psi (x^{\sl 0},\underline{x};X) \midd \al \rangle
= {\sqrt Z_2}\, \langle \be \midd \psi_{\mbox{\ss out/in}} (x^{\sl 0},\underline{x};X) \midd \al \rangle
\eeq
holds as usual in the mean only we obtain the generalized LSZ reduction formula for Dirac matter in an analogous way as for scalar matter
\beqq \label{149}
& & \langle h_1\dots h_m; h'_1\dots h'_{m'}\, \mbox{out}\midd k_1\dots k_n;  k'_1\dots k'_{n'}\,\mbox{in} \rangle = \nonumber \\
& &\quad\quad\quad (-)^{m+n}\,
\Pi_{i = 1}^n \lim_{K_i \ar k_i} \Pi_{r = 1}^{n'} \lim_{K'_r \ar k'_r}
\Pi_{j = 1}^m \lim_{H_j \ar h_j} \Pi_{s = 1}^{m'} \lim_{H'_s \ar h'_s}
\nonumber \\
& &\quad \left( \frac{i}{\sqrt Z_2} \right)^{n+n'}\!\!\!\!\!\!\!\!\!\!
\Pi_{i = 1}^n \intx_i \intX\!_i\, \La^{-4 n}\,
\Pi_{r = 1}^{n'} \intx'_r \intX\!'\!_r\, \La^{-4 m} \nonumber \\
& &\quad \left( \frac{i}{\sqrt Z_2} \right)^{m+m'}\!\!\!\!\!\!\!\!\!\!\!\!
\Pi_{j = 1}^m \inty_j \intY\!_j\, \La^{-4 m}\,
\Pi_{s = 1}^{m'} \inty'_s \intY\!'\!_s\, \La^{-4 n} \nonumber \\
& &\quad\quad\quad\quad\quad\quad
\Pi_{j = 1}^m \bar u (h_j,t_j)\, e^{ih_j y_j + iH_j Y_j}
\Big( i{\dsl\rvec}_j - m \Big) \\
& &\quad\quad\quad\quad\quad\quad
\Pi_{r = 1}^{n'} \bar v (k'_r,s'_r)\, e^{-ik'_r x'_r - iK'_r X'_r}
\Big( i{\dsl\rvec}'_r - m \Big) \nonumber \\
& &\quad \langle 0 \midd
T\Big( \dots \bar\psi (y'_s,Y'_s) \dots \psi (y_j,Y_j) \dots
\bar\psi (x_i,X_i) \dots \psi (x'_r,X'_r) \dots \Big)
\midd 0 \rangle \nonumber \\
& &\quad\quad\quad\quad\quad\quad
\Pi_{i = 1}^n \Big( -i{\dsl\lvec}_i - m \Big)
u (k_i,s_i)\, e^{-ik_i x_i - iK_i X_i} 
\nonumber \\
& &\quad\quad\quad\quad\quad\quad
\Pi_{s = 1}^{m'} \Big( -i{\dsl\lvec}'_s - m \Big)
v (h'_s,t'_s)\, e^{ih'_s y'_s + iH'_s Y'_s}
\nonumber
\eeqq
linking general matrix elements of the gravitational $S$-operator to the gravitational limit of vacuum expectation values of time-ordered products of interacting Dirac field operators.

\section{Generalized LSZ Reduction Formulae: Gauge Field}
In this section we establish the LSZ reduction formulae for the gauge field thereby expressing $S$-matrix elements with external gauge particles as the gravitational limit of truncated Fourier-transformed vacuum expectation values of time-ordered products of field operators of the interacting theory on the mass shell. Note that the generating functional of the latter has been defined and renormalized to one-loop which puts us in a position to perturbatively calculate any $S$-matrix element of interest in quantum gravity \cite{chw4}.

Let us start with the field equation Eqn.(\ref{74}) with $\la = 1$ for the interacting transversal gauge field
\beq \label{150}
\left( -\pa^2 - \mu^2\right) A^T_\mu\,^\al(x,X) = J_\mu\,^\al(x,X).
\eeq
Here $J_\mu\,^\al$ contains all the interaction terms including possible self-interact-ions. We deal with the transversal part of the gauge field as it is for this field components only that a sensible asymptotic condition can be formulated \cite{cli}. Note that the case $\la \neq 1$ could be dealt with in the usual way.

We next define the $in$- and $out$-fields $A^T_\mu\,^\al_{\mbox{\ss in/out}}$
\beqq \label{151}
& & {\sqrt Z_3}\, A^T_\mu\,^\al_{\mbox{\ss in/out}}(x,X) = A^T_\mu\,^\al(x,X) \\
& &\quad - \inty\intY \La^{-4}\, \Del_{\mu\nu}\,^{\al\be}_{\mbox{\ss ret/adv}} (x-y, X-Y)\, J^\nu\,_\be(y,Y) \nonumber
\eeqq
where $\Del_{\mu\nu}\,^{\al\be}_{\mbox{\ss ret/adv}}$ are the retarded and advanced Green functions transversal in inner space
\beq \label{152}
\left(-\pa^2 - \mu^2\right) \Del_{\mu\nu}\,^{\al\be}_{\mbox{\ss ret/adv}}
(x-y, X-Y) = \eta_{\mu\nu}\, ^T\!\de^{\al\be}(X-Y)\, \de^4(x-y)
\eeq 
which vanish for $x^{\sl 0} < y^{\sl 0}$ and $x^{\sl 0} > y^{\sl 0}$ respectively.

By construction $A^T_\mu\,^\al_{\mbox{\ss in/out}}$ fulfil the field equations Eqn.(\ref{74}) for $J = 0$
\beq \label{153}
\left(-\pa^2 - \mu^2\right) {\sqrt Z_3}\, A^T_\mu\,^\al_{\mbox{\ss in/out}}(x,X) = 0
\eeq
for the transversal part of the free gauge field obeying
\beq \label{154}
\pa^\mu A^T_\mu\,^\al_{\mbox{\ss in/out}}(x,X) = 0.
\eeq
We have quantized this field which only contains physical degrees of freedom in section 4. It is given in terms of creation and annihilation operators and the two polarization vectors as
\beqq \label{155}
& & A^T_\mu\,^\al_{\mbox{\ss in/out}} (x,X) = \intko \intK\, \La^4\,
\sum_{\ga = 1}^3\, \sum_{\Ga = 1}^3 \\
& &\quad \ep_\mu (k,\ga)\, \Ep^\al (K,\Ga)
\left\{ a_{+\, \mbox{\ss in/out}} (\underline{k},\ga;K,\Ga)\, e^{-ikx - iKX} + \mbox{h.c.} \right\}.
\nonumber
\eeqq 

As noted above the asymptotic condition is formulated for the transversal part of the gauge field $A^T_\rho\,^\al(x,X) \ar {\sqrt Z_3}\, A^T_\rho\,^\al_{\mbox{\ss in/out}}(x,X)$ for $x^{\sl 0}\ar \pm \infty$ and holds in the mean only
\beq \label{156}
\lim_{x^{\sl 0} \ar \pm\infty} \plangle \be \midd
A^T_\rho\,^\al (x^{\sl 0},\underline{x};X) \midd \al \prangle
= {\sqrt Z_3}\, \plangle \be \midd A^T_\rho\,^\al_{\mbox{\ss out/in}} (x^{\sl 0},\underline{x};X) \midd \al \prangle.
\eeq
In the sequel $\plangle\dots \midd$ and $\midd \dots \prangle$ denote physical gauge particle states containing only transversal space-time polarizations.

Starting with a general $S$-matrix element $\plangle \be\, \mbox{out}\midd k\, \al\, \mbox{in} \prangle$, where the $in$-state contains a one-particle state with momentum $k$, space-time and inner polarizations $\ga$ and $\Ga$ respectively, we again perform the first reduction in some detail
\beqq \label{157}
S_{\be,k \al} &=& \plangle \be\, \mbox{out}\midd k\, \al\, \mbox{in}
\prangle
= \plangle \be\, \mbox{out}\midd \bar a^\dagger_{\mbox{\ss in}} (\underline{k},\ga,\Ga) \midd \al\, \mbox{in} \prangle \nonumber \\
&=& \lim_{K \ar k} \plangle \be\, \mbox{out}\midd
a^\dagger_{+\, \mbox{\ss in}} (\underline{k},\ga;K,\Ga)
\midd \al\, \mbox{in} \prangle \nonumber \\
&=& \plangle \be\, -\!k\, \mbox{out}\midd \al\, \mbox{in} \prangle
\nonumber \\
& &+ \lim_{K \ar k} \plangle \be\, \mbox{out}\midd
a^\dagger_{+\, \mbox{\ss in}} (\underline{k},\ga;K,\Ga)
- a^\dagger_{+\, \mbox{\ss out}} (\underline{k},\ga;K,\Ga) \midd \al\, \mbox{in} \prangle \nonumber \\
&=& \mbox{elastic contribution} \nonumber \\
& &- \lim_{K \ar k}\, i\,
\int\! d^{\sl 3}x \intX\La^{-4}\, e^{-ikx - iKX}
\ep^\rho (k,\ga) \Ep_\al (K,\Ga) \nonumber \\
& &\quad\quad\quad  \pa_{\sl 0}\lrvec\,\, \plangle \be\, \mbox{out}\midd
A^T_\rho\,^\al_{\mbox{\ss in}} (x^{\sl 0},\underline{x};X) -
A^T_\rho\,^\al_{\mbox{\ss out}} (x^{\sl 0},\underline{x};X) \midd \al\, \mbox{in} \prangle \nonumber \\
&=& \mbox{elastic contribution} \nonumber \\
& &+ \lim_{K \ar k}\, \frac{i}{\sqrt Z_3}\,
\Big(\lim_{x^{\sl 0} \ar \infty} - \lim_{x^{\sl 0} \ar -\infty}\Big)
\int\! d^{\sl 3}x \intX\La^{-4}\, e^{-ikx - iKX}\nonumber \\
& &\quad\quad\quad  \ep^\rho (k,\ga) \Ep_\al (K,\Ga)
\pa_{\sl 0}\lrvec\,\, \plangle \be\, \mbox{out}\midd
A^T_\rho\,^\al (x^{\sl 0},\underline{x};X) \midd \al\, \mbox{in} \prangle \nonumber \\
&=& \mbox{elastic contribution} \nonumber \\
& &+ \lim_{K \ar k}\, \frac{i}{\sqrt Z_3}\,
\intx\, \frac{\pa}{\pa x^{\sl 0}} \intX\La^{-4}\, e^{-ikx - iKX} \nonumber \\
& &\quad\quad\quad \ep^\rho (k,\ga) \Ep_\al (K,\Ga)
\pa_{\sl 0}\lrvec\,\, \plangle \be\, \mbox{out}\midd
A^T_\rho\,^\al (x^{\sl 0},\underline{x};X) \midd \al\, \mbox{in} \prangle \\
&=& \mbox{elastic contribution} \nonumber \\
& &+ \lim_{K \ar k}\, \frac{i}{\sqrt Z_3}\,
\intx\, \intX\La^{-4}\, e^{-ikx - iKX} \ep^\rho (k,\ga) \Ep_\al (K,\Ga) \nonumber \\
& &\quad\quad\quad 
\Big( {\pa\rvec}_{\sl 0}^2 - {\pa\lvec}_{\sl 0}^2 \Big)
\plangle \be\, \mbox{out}\midd A^T_\rho\,^\al (x^{\sl 0},\underline{x};X)
\midd \al\, \mbox{in} \prangle \nonumber \\
&=& \mbox{elastic contribution} \nonumber \\
& &+ \lim_{K \ar k}\, \frac{i}{\sqrt Z_3}\,
\intx\, \intX\La^{-4}\, e^{-ikx - iKX} \ep^\rho (k,\ga) \Ep_\al (K,\Ga) \nonumber \\
& &\quad\quad\quad 
\Big( {\pa\rvec}^2 + \mu^2 \Big)
\plangle \be\, \mbox{out}\midd A^T_\rho\,^\al (x;X)
\midd \al\, \mbox{in} \prangle \nonumber \\
&=& \mbox{elastic contribution} \nonumber \\
& &+ \lim_{K \ar k}\, \frac{i}{\sqrt Z_3}\,
\intx\, \intX\La^{-4}\, e^{-ikx - iKX} \ep^\rho (k,\ga) \Ep_\al (K,\Ga) \nonumber \\
& &\quad\quad\quad 
\Big( {\pa\rvec}^2 + \mu^2 \Big)
\plangle \be\, \mbox{out}\midd A_\rho\,^\al (x;X)
\midd \al\, \mbox{in} \prangle \nonumber \\
&=& \mbox{elastic contribution} \nonumber \\
& &+ \lim_{K \ar k}\, \frac{i}{\sqrt Z_3}\,
\intx\, \intX\La^{-4}\, e^{-ikx - iKX} \nonumber \\
& &\quad\quad\quad \ep^\rho (k,\ga) \Ep_\al (K,\Ga) 
\plangle \be\, \mbox{out}\midd J_\rho\,^\al (x;X)
\midd \al\, \mbox{in} \prangle, \nonumber
\eeqq
where we have made crucial use of the transversality restricting $\ga$ to $\ga = 1,2,3$, the asymptotic condition Eqn.(\ref{156}) and the on-shell condition $k^2 = \mu^2$.

Iterating the above steps we finally obtain the generalized LSZ reduction formula for the gauge field
\beqq \label{158}
& &\langle h_1\dots h_m\, \mbox{out}\midd k_1\dots k_n\, \mbox{in} \rangle
= \left( \frac{i}{\sqrt Z_3} \right)^{m+n} \Pi_{i = 1}^n \lim_{K_i \ar k_i} \Pi_{j = 1}^m \lim_{H_j \ar h_j} \nonumber \\
& &\quad\quad \Pi_{i = 1}^n \intx_i \intX\!_i\, \La^{-4 n}\,
\Pi_{j = 1}^m \inty_j \intY\!\!_j\, \La^{-4 m} \nonumber \\
& &\quad\quad\quad \Pi_{i = 1}^n \ep^{\rho_i} (k_i,\ga_i)
\Ep_{\al_i} (K_i,\Ga_i)
e^{-ik_i x_i - iK_i X_i} \Big( {\pa\rvec}_i^2 + \mu^2 \Big) \\
& & \quad\quad \plangle 0 \midd
T\Big(\dots A_{\rho_i}\,\!^{\al_i} (x_i,X_i)\dots
A_{\si_j}\,\!^{\be_j} (y_j,Y_j) \dots \Big) \midd 0 \prangle \nonumber \\
& &\quad\quad\quad \Pi_{j = 1}^m \Big( {\pa\lvec}_j^2 + \mu^2 \Big)
e^{ih_j y_j + iH_j Y_j} \ep^{\si_j} (h_j,\ga'_j)
\Ep_{\be_j} (H_j,\Ga'_j). \nonumber
\eeqq

\section{Conclusions}
In this paper we have first reviewed the construction of the gauge theory of volume-preservering diffeomorphisms. Key to that construction is that both matter and gauge fields carry an additional continuous number of inner degrees of freedom allowing to represent the infinite-dimensional gauge group in field space. As the actions of the matter and gauge fields are by construction invariant under both Poincar\'e and volume-preserving diffeomorphism group transformations Noether's theorem yields two independently conserved four-vectors - inertial and gravitational energy-momentum which in our approach play physically fundamentally different roles \cite{chw2}.

In order to further uncover the physical content of the theory we then have canonically quantized the free scalar and Dirac fields properly dealing with the additional degrees of freedom in the quantization process. By construction both types of free fields do not only carry the usual inertial, but also gravitational energy-momentum for which we get explicit expressions in terms of creation and annihilation operators. Using those operators we then have constructed the respective Fock spaces of stationary states which are labelled by the usual quantum numbers and - in addition - the gravitational energy-momentum four-vector. Finally we have introduced what we call the "gravitational limit". It is defined by collapsing the gravitational energy-momentum onto the inertial energy-momentum to account for their observed equality. Mapping a product of two Minkowski spaces onto one yields a singular limit and requires in the case of commutation relations of creation and annihilation operators a regularization of the volume of inner $X$-space. Using the freedom of setting that regularized volume equal to the hitherto unspecified $\La^4$ allows us to demonstrate that the commutation relations in the gravitational limit are equivalent to those for a free quantum field without additional inner degrees of freedom - which corresponds to an observable particle. As a result we have a description of observable particles which "remember" that they carry both inertial and gravitational energy-momentum - one allowing to express energy-momentum conservation, the other being the current to which the gauge field couples.

In order to properly quantize the gauge field we have employed Stueckelberg's approach adding both a gauge fixing and a mass term to the original gauge field Lagrangian \cite{cli}. This has allowed us to quantize the gauge field as a set of scalar fields. As the corresponding commutation relations come along with both signs the Fock space of stationary states has no definite metric and has for the observable states to be restricted to positive norm states in the usual way. The corresponding states are transversal in relation to the inertial energy-momentum vector $k$. The one new key feature related to the additional inner degrees of freedom is the occurrence of additional inner polarization degrees of freedom. As the gauge fields are divergence-free in inner space field quanta can carry only transversal inner polarization. Hence, a physical field quantum carries two independent polarizations being spin-1 in both space-time and inner space. Taking the gravitational limit finally leaves us with a physical Fock space of stationary states labelled by the usual momentum vector and a tensor-product of two transversal spin-1 polarizations - or with transversal spin-2 field quanta which describe the observable exchange particles for the gravitational force in our approach.

Having constructed the asymptotic states of the theory we then have turned to the definition of the $S$-matrix in quantum gravity. Our starting point have been the transition amplitudes of asymptotic $in$- to $out$-states which carry both inertial and gravitational energy-momentum. The corresponding "large" $S$-matrix displays the usual properties, most importantly it is unitary. The $S$-matrix in quantum gravity or "small" $S$-matrix is then defined as the gravitational limit of the transition amplitudes of asymptotic $in$- to $out$-states and relates $in$- and $out$-states of observable particles carrying gravitational equal to inertial energy-momentum. Finally we have shown that the small $S$-matrix enjoys the same properties as the large one, notably it is unitary if the large one is.

In the last two sections of this paper we have established generalized LSZ reduction formulae for scalar and Dirac matter and the gauge field which allow to express $S$-matrix elements as the gravitational limit of truncated Fourier-transformed vacuum expectation values of time-ordered products of field operators of the interacting theory. Together with the generating functional of the latter established in \cite{chw4} we can in principle compute any transition amplitude to any order in perturbative quantum gravity. Such calculations yielding well-defined transition amplitudes are obviously only meaningful if both divergent space-time and inner space integrals appearing in the theory are renormalizable and regularizable in a way respecting the theory's symmetries respectively.

In \cite{chw4} we have given a few hints at the theory's renormalizability, however no proof - the absence of which provides ample ground for scepticism vis-a-vis our approach. In the meantime based on a BRST-type of symmetry given in \cite{chw4} we have established a full proof of the renormalizabilty w.r.t divergent space-time integrals and consistent regularizability w.r.t. divergent inner space integrals of the gauge theory of volume-preserving diffeomorphisms \cite{chw5}.

We note that at any stage of the calculation of $S$-matrix elements before projecting the gravitational momentum of an asymptotic - and hence free and observable - state onto the inertial momentum of this state the gravitational momentum is a mathematical entity carried separately through the interacting theory like colour is carried through interacting QCD - so the interacting theory and its correlation functions are always defined on both the $x$- and $X$-space as shown e.g. in Eqn.(\ref{143}). It is only for asymptotic, observable states that the projections are done - and these projections have explicitly been performed in sections 3 and 4 without mathematical difficulties beyond the ones discussed.

Finally why and on what basis do we claim that our approach might serve as a quantum theory of gravity?

On the one hand it fulfils general requirements such as the universal coupling to all other fields \cite{chw2,chw4}. On the other if it is a theory of gravity we should be able to analyse physical situations for which we can make and compare predictions both within the framework of the theory presented as well as within at least the standard framework of Newtonian gravity dealt with quantum-mechanically. 

With the results of this paper we can evaluate the gauge theory of volume-preserving diffeomorphisms perturbatively for the scattering of particles and one situation of physical interest is the gravitational scattering of two particles with different masses. Having done this in \cite{chw5} we find that the gravitational scattering cross-section of two Dirac particles with different masses indeed reduces in an appropriate limit to the cross-section of a non-relativistic particle scattering off an infinitely heavy scatterer calculated quantum mechanically in Newtonian gravity.

This is an important indication that the gauge field theory of volume-preserving diffeomorphisms can be viewed as a renormalizable quantum theory of gravity.

What remains as a difficult open question, however, is the classical limit to the quantum theory presented. To work out what happens for $\hbar \ar 0$ is indeed even less trivial than for other quantum field theories due to the fact that $\hbar$ appears in the theory not only in the usual form, but in addition through the Planck scale which is the length scale $\La$ appearing in the theory - and the Planck scale itself vanishes with $\hbar \ar 0$ as well. We have worked out the leading order to limit $\hbar \ar 0$ in \cite{chw3} which results in a linear relativistic theory of gravitation. Its non-relativistic limit yields Newtonian gravitation without further assumptions - however, for consistency reasons the full classical limit has to contain non-linearities which will be very hard to work out directly.

As the absence of such a direct comparison to Einstein's theory provides ample ground for scepticism vis-a-vis our approach we have established a strong argument on the basis of symmetry considerations that in the low energy approximation General Relativity comes indeed to the fore as the classical limit to the gauge theory of volume-preserving diffeomorphisms \cite{chw7}.

\section{Notations and Conventions}

Generally, ({\bf M}$^{\sl 4}$,\,$\eta$) denotes the four-dimensional Minkowski space with metric $\eta=\mbox{diag}(1,-1,-1,-1)$, small letters denote space-time coordinates and parameters and capital letters denote coordinates and parameters in inner space.

Specifically, $x^\la,y^\mu,z^\nu,\dots\,$ denote Cartesian space-time coordinates. The small Greek indices $\la,\mu,\nu,\dots$ from the middle of the Greek alphabet run over $\sl{0,1,2,3}$. They are raised and lowered with $\eta$, i.e. $x_\mu=\eta_{\mu\nu}\, x^\nu$ etc. and transform covariantly w.r.t. the Lorentz group $SO(\sl{1,3})$. Partial differentiation w.r.t to $x^\mu$ is denoted by $\pa_\mu \equiv \frac{\pa\,\,\,}{\pa x^\mu}$. Small Latin indices $i,j,k,\dots$ generally run over the three spatial coordinates $\sl{1,2,3}$ \cite{cli,bjd}.

$X^\al, Y^\be, Z^\ga,\dots\,$ denote inner coordinates and $g_{\al\be}$ the flat metric in inner space with signature $+,-,-,-$. The metric transforms as a contravariant tensor of Rank 2 w.r.t. ${\overline{DIFF}}\,{\bf M}^{\sl 4}$. Because Riem$(g) = 0$ we can always globally choose Cartesian coordinates and the Minkowski metric $\eta$ which amounts to partially fixing the gauge to Minkowskian gauges. The small Greek indices $\al,\be,\ga,\dots$ from the beginning of the Greek alphabet run again over $\sl{0,1,2,3}$. They are raised and lowered with $g$, i.e. $x_\al=g_{\al\be}\, x^\be$ etc. and transform as vector indices w.r.t. ${\overline{DIFF}}\,{\bf M}^{\sl 4}$. Partial differentiation w.r.t to $X^\al$ is denoted by $\nabla_\al \equiv \frac{\pa\,\,\,}{\pa X^\al}$. 

The same lower and upper indices are summed unless indicated otherwise.

\end{document}